# Wettability and $sp^2/sp^3$ ratio effects on supercapacitor performance of N-doped hydrogenated amorphous Carbon Nanofoam

Subrata Ghosh[1,2*], Giacomo Pagani[1], Andrea Macrelli[1], Alberto Calloni[3], Gianlorenzo Bussetti[3], Andrea Lucotti[4], Matteo Tommasini[4], Raffaella Suriano[5], Marco Agozzino[1], Giorgio Divitini[6], Yurii P. Ivanov[6], Veronica Piazza[7], Valeria Russo[1], Agnieszka M. Jastrzębska[2], Cinzia Casiraghi[8], Andrea Li Bassi[1], Carlo S. Casari[1*]

[1] *Micro and Nanostructured Materials Laboratory – NanoLab, Department of Energy, Politecnico di Milano, via Lambruschini, Milano, 20156, Italy*
[2] *Faculty of Mechatronics, Warsaw University of Technology, sw. Andrzeja Boboli 8, 02-525 Warsaw, Poland*
[2] *Solid Liquid Interface Nano-Microscopy and Spectroscopy (SoLINano-Σ) lab, Department of Physics, Politecnico di Milano, Piazza Leonardo da Vinci 32, 20133 Milano, Italy*
[4] *Department of Chemistry, Materials and Chemical Engineering, "Giulio Natta" Politecnico di Milano, Milano, 20133, Italy*
[5] *Laboratory of Chemistry and Characterization of Innovative Polymers (ChIPLab), Department of Chemistry, Materials and Chemical Engineering "Giulio Natta", Politecnico di Milano, Piazza Leonardo da Vinci 32, 20133 Milano, Italy*
[6] *Electron Spectroscopy and Nanoscopy, Istituto Italiano di Tecnologia, via Morego 30, 16163 Genova, Italy*
[7] *Laboratory of Catalysis and Catalytic Processes, Dipartimento di Energia, Politecnico di Milano, via La Masa 34, 20156 Milano, Italy*
[8] *Department of Chemistry, University of Manchester, Manchester, M13 9PL, UK*

**Abstract**: Pulsed laser-deposited amorphous carbon nanofoams are potential candidate for electrochemical energy storage applications due to ultra-light weight, large volumetric void fractions, and co-existence of *sp*, $sp^2$ and $sp^3$ carbon hybridization. It is known that charge-storage in carbon nanostructures containing disordered $sp^2$-domains is determined by their wettability, surface area, and porosity. However, their charge-storage performance is limited to the areal capacitance of the order of a few mF/cm². We enhanced the supercapacitor performance of nitrogen-doped amorphous carbon nanofoam by engineering its wettability and $sp^2$-C/$sp^3$-C ratio by vacuum annealing. The specific capacitance was enhanced by about fifty times and the device voltage increased from 0.8 to 1.1 V compared to as-grown carbon nanofoam. In addition, we examined for the first time the initial increase in specific capacitance of the aqueous symmetric supercapacitor with respect to the scan rate, employing *in-situ* measurements coupling Raman spectroscopy and electrochemistry. We attribute this effect, observed but generally not explained in previous works in the literature, to the electrochemical activation induced by structural changes during the charge storage performance. This optimization of pulsed laser-deposited carbon nanofoam may open an avenue for fabricating lightweight and porous nanostructures for advanced macro-to-micro-supercapacitor devices.

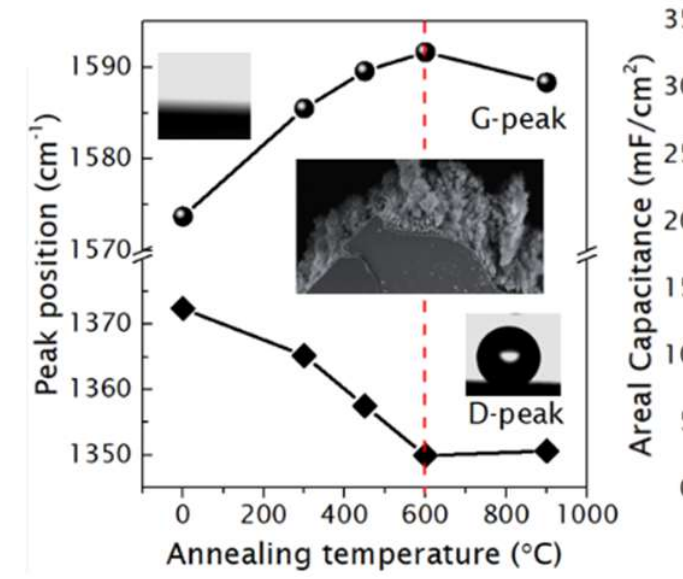
1590
1580
1570
1370
1360
1350
Peak position (cm⁻¹)
G-peak
D-peak
0
200
400
600
800
1000
Annealing temperature (°C)

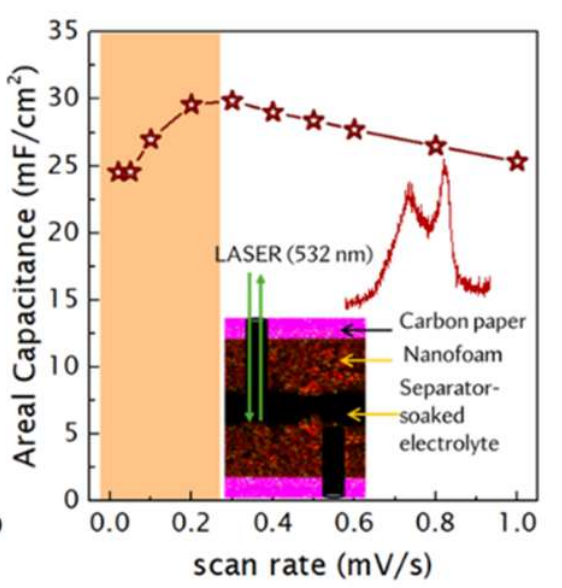
35
30
25
20
15
10
5
0
Areal Capacitance (mF/cm²)
LASER (532 nm)
Carbon paper
Nanofoam
Separator-
soaked
electrolyte
0.0
0.2
0.4
0.6
0.8
1.0
scan rate (mV/s)

*Keywords: Nanocarbons; wettability; graphitic quality; in-situ spectro-electrochemistry; supercapacitor*

Corresponding author email: subrata.ghosh@pw.edu.pl (S.G.) and carlo.casari@polimi.it (C.S.C.)

## 1. Introduction

Supercapacitors have been developed into viable applications and are anticipated to be a key class of electrochemical energy storage devices for the clean energy transition owing to their excellent power density, cycle life, and capability to work at extreme conditions (e.g. -40 °C to 80 °C).[1–3] One of the biggest challenges in the development of this technology for both the academic and industrial communities is to enhance the energy density in order to compete with batteries. The crucial research focuses are the development of advanced electrode materials design, the investigation of the best electrolytes, as well as the study of their best combination.[4–7]

Among electrode materials, carbon-based nanostructures are promising as they offer excellent structural quality, tuneable porosity, and the possibility to be doped or decorated with other systems to form nanostructures.[8] The pure $sp^2$-bonded carbon, graphene, unfortunately, failed to reach its theoretical capacitance (550 F/g) despite a high surface area of 2630 g/cm² and excellent structural quality.[9] On the other hand, the $sp^3$-bonded carbon allotrope, diamond, is well-known to provide excellent electrochemical stability over the widened potential window. However, it also shows poor charge-storage capacity as the main drawback for use in supercapacitors.[10][11] Recent studies claim nanoporous carbons containing small graphene ($sp^2$-hybridized) flakes with high structural disorder,[12] and pores smaller than one nanometer[13] are promising as high-performance supercapacitor electrodes. Factors influencing the performance of carbon-based supercapacitors or electric double-layer capacitors are wettability, structural quality, porosity, surface area, type of hybridization (*sp*, $sp^2$ and $sp^3$), and heteroatom doping.[1,2,5,14–16] In carbon nanostructures such as carbon nanotubes (CNTs), vertical graphenes etc, the hydrophobic nature prevents the electrode/electrolyte interaction. Hence, tuning the wettability by heteroatom such as boron, nitrogen, and sulphur doping or oxygen functionalization can make them hydrophilic to improve the charge-storage performance.[5][7][15] Doping and functionalization also improve the charge-storage/transfer kinetics, contribute to pseudocapacitance, and increase the double-layer

capacitance.[5][17][18] Among potential dopants, nitrogen is a promising choice as it provides more electroactive sites, introduces more disorder in the structure, improves the wettability and hence better capacitive performance compared to the undoped one.[19] Machine learning-assisted design of porous carbon nanostructures containing nitrogen and oxygen functional groups has been anticipated as promising to design optimized supercapacitor electrodes.[14] However, it has been reported that reduced graphene oxide (RGO) with specifically tuned carbon/oxygen ratio and electrical conductivity, when annealed at 200 °C exhibits higher specific capacitance than when pristine or when annealed at more than 200 °C.[20] In the report on boron and nitrogen co-doped porous carbon, thermal annealing of the co-doped porous carbon at 800 °C delivers higher surface area, outstanding ion diffusion and hence excellent supercapacitor performance compared to the co-doped porous carbon thermally annealed at 900 °C.[18] Unfortunately, toxic and unsafe precursors such as $NH_3$[19] and urea[18] were widely used to synthesize N-doped porous carbons. Moreover, it should be noted that the optimization the dopant content and specific bonding arrangement is highly crucial to creating desirable pore layout and specific capacitance.[5][21][22] This result suggests that the doped-carbon nanostructures with desired morphological and optimized structural properties are decisive in obtaining high charge-storage performance when used as supercapacitor electrode. Moreover, the active electrode materials are generally mixed with a binder and conductive agent to produce a supercapacitor electrode,[17] where binders and conductive agents do not contribute to charge storage. Instead, these additives prevent the electrolyte ions from accessing the surface.[8] We also highlight that one can find reports on carbon nanostructures or other nanostructures where a material with lower thickness or low mass loading delivered higher specific capacitance.[23] Fabricated thick films of nanostructures often present a limited accessible surface area; longer diffusion pathways prevent ion transport and therefore lower charge-storage performance is achieved. Thus, binder-free and conductive-agent-free ultrathick porous electrodes are of scientific and industrial interest.[24–29] However, the right supercapacitive material muse be selected, in particular among pseudocapacitive materials.[6][30] Amorphous carbons consist of *sp*, $sp^2$ and $sp^3$ hybridizations, with plenty of electrochemically active sites and local structural disorder that can be utilized as an energy storage electrode. However, poor electrical conductivity limits its use as an electrode.[31] A free-standing amorphous carbon film was also prepared by chemical vapor deposition, which was transferred from the Ni substrate to the desired one to fabricate a supercapacitor device, but the areal capacitance was limited to 0.28 mF/cm$^2$ at 50 mV/s.[32] Our previous investigations show that an aqueous supercapacitor composed of self-standing amorphous carbon foam, which is lightweight and more than 90% porous, delivers the areal capacitance of ~1 mF/cm$^2$ at 0.1 V/s, in spite of being hydrophilic and highly disordered.[33]

In the present work, a hydrogenated amorphous carbon nanofoam containing nitrogen was synthesized directly on the current collector by pulsed laser deposition (PLD) under $N_2$ background gas at room temperature, avoiding any use of binders and conductive agents. The wettability, $sp^2$-C/ $sp^3$-C ratio, and nitrogen content of the nanofoams were tuned by vacuum annealing at different temperatures. Finally, an aqueous symmetric supercapacitor device was assembled to investigate the charge storage performance of those nanofoams and correlate them with their physico-chemical characteristics. While investigating the charge-storage performance, we observed an initial increase of capacitance with respect to the scan rate, which is rare but observed in previous reports on nanoporous gold/$MnO_2$ hybrid electrode[34], $MnO_2$ nanorods on carbon nanotubes,[35] activated N-doped amorphous carbon nanofoams,[36] and metal-free MXene[37]. However, no explanation was provided for the enhancement. There may be electrochemical activation-induced structural changes during the charge/discharge process, which is often observed at the initial cycles of prolonged charge-discharge tests.[6]Therefore, *in-situ* Raman spectro-electrochemistry of the prototype aqueous symmetric supercapacitor was performed and the changes in graphitic quality due to the electrochemical activation are attributed to the initial increase in specific capacitance with respect to the scan rate. This result shows that nanofoam with optimized properties can be a potential electrode for binder-free thin-film supercapacitor applications.

## 1. Experimental methods

### 1.1. *Synthesis of carbon nanofoam*

Binder-free carbon nanofoams were directly grown on the carbon paper and Si substrate using PLD at room temperature (Scheme 1a). More details of carbon nanofoam synthesis were reported in our previous work. [38] A graphite target of 1-inch was placed on translating and rotating target holder and the ablation was carried out using a Nd:YAG pulsed nanosecond-laser (2$^{nd}$ harmonic at 532 nm, pulse duration 5-7ns, repetition rate 10 Hz). Two-step deposition was carried out: a buffer layer was formed at the deposition pressure of 5 Pa for 2 min, followed by the deposition of a porous structure for 30 min at 300 Pa under the mixture of 95% nitrogen and 5% hydrogen background gas and at a laser pulse energy of about 416 mJ with a fluence of 6.5 J/cm$^2$. Prior to the deposition, the chamber was evacuated down to $10^{-3}$ Pa as a turbomolecular pump connected to a rotary pump. The substrate was placed on rotating substrate holder at a distance of 40 mm from the target.

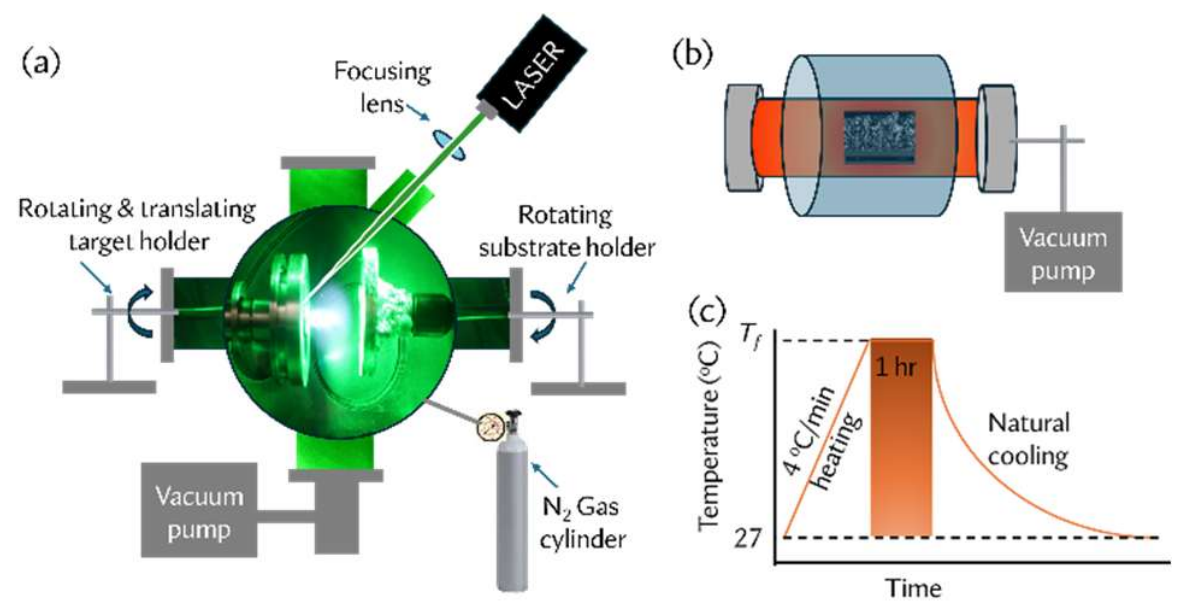

Scheme 1: Schematic of (a) pulsed laser deposition, (b) vacuum annealing and (c) corresponding temperature *vs* time profile for the deposition of N-doped hydrogenated amorphous carbon nanofoam.

As-deposited nanofoams were treated at temperatures of 300, 450, 600 and 900 °C in a tube furnace for 1 hour under vacuum to modify the $sp^2$-C/ $sp^3$-C ratio, dopant environment and functionalities (Scheme 1b). The ramping rate to reach the desired temperature was 4 °C/min (Scheme 1c) and the pressure during the annealing was maintained at $5\times10^{-3}$ Pa by rotary and turbomolecular pump. For the functionalization, the optimized carbon nanofoam was dipped into 8M $HNO_3$ solution for 7 hours 30 minutes hours at the temperature of 120 °C, then it was washed several times with deionized water until the pH of the solution reached 6, and finally it was dried in an oven at about 90 °C.[39]

### 1.2. *Morphological and structural characterization.*

A field-emission scanning electron microscope (FE-SEM, ZEISS SUPRA 40, Jena, Germany) was employed to investigate the morphology of the composite, where an *in-lens* detector operates in a high vacuum. Energy dispersive X-ray spectra (EDXS) were recorded at a 20 kV acceleration voltage, using AZtec software for elemental analysis, volumetric void fraction estimation, and mass density. The instrument was equipped with a Peltier-cooled silicon drift detector (Oxford Instruments). High-resolution Transmission electron micrographs were acquired on an image-corrected Thermo Fisher Spectra 300 operated at 80 kV, using a Ceta CMOS camera.

Thermogravimetric analysis (TGA) was carried out in a Hitachi STA7300 instrument, following a heating rate of 4 °C/min from room temperature to 1050 °C under He inert gas (100 NmL/min) . The evolution of gas species was monitored in an online mass spectrometer (HPR-20 EGA, Hidden Analytical). For TGA, the pristine CFs were scratched from the Si substrate and the powder was collected.

X-ray photoelectron spectroscopy (XPS) was utilized to investigate the elemental composition and bonding environment of the pristine and annealed carbon nanofoams. XPS data were collected using a non-monochromated X-ray source using a Mg anode (photon energy 1253.6 eV), maintained at a power of 200 W, and pressure of ~$1x10^{-10}$ torr. The kinetic energy of the photoemitted electrons was recorded using a hemispherical analyser with a 150 mm mean radius, PHOIBOS150 from SPECS GmbH. The spectra were acquired with a pass energy of 20 eV, with an energy resolution of 0.9 eV (full-width half maximum, FWHM). Peaks were fitted after Shirley background subtraction using CasaXPS software, and at.% elemental compositions were extracted from peak area ratios after correction by Scofield relative sensitivity factors (C=1.0, N=1.77, O=2.85).[40] For C1*s*, the asymmetric $sp^2$–C peak (calibrated at 284.8 eV) was fitted with Gaussian-Lorentzian lineshape (GL(30)) with asymmetric factor (T200) and other symmetric carbon peaks with GL(30) by setting the range of FWHM to 1.2–2 eV. The FWHM of oxygenated carbon peaks, deconvoluted O1s peaks, and deconvoluted N1*s* peaks were set to 1.8–2.2 eV. The $sp^3$–C peak was generally shifted by 0.7–1 eV from $sp^2$–C and hydroxyl/ether, carbonyl, and carboxylic groups were shifted approximately 1.5, 3, and 4.5 eV higher, respectively.

Micro-Fourier Transform Infrared (FT-IR) measurements on the carbon foam deposited on silicon were performed using the Nicolet Nexus interferometer coupled with a Thermo-Nicolet Continuum infrared microscope and a cooled MCT detector (77 K). The films deposited on silicon were measured in transmission through two co-axial Cassegrain 15× infrared objectives (one for focalizing the IR beam, the other for the collection of the transmitted light). The analyzed area was approximately 200 µm x 200 µm.

The Raman spectra of all samples were recorded by Renishaw *in-via* Raman spectrometer, UK. The spectra were acquired using a 514.5 nm laser with a power of 0.4 mW on the sample, a 1800 line/mm grating spectrometer, a 50× objective lens, and 20 summed spectra (10 s each).

### 2.3. *Wettability measurement*

Wettability measurements were performed using an OCA 15plus instrument (Dataphysics Co., Filderstadt, Germany), equipped with a CCD camera to capture side views of drop images, and a 500 µL Hamilton syringe to dispense water droplets. Water for chromatography (LC-MS Grade, LiChrosolv®) supplied by Merck (KGaA, Darmstadt, Germany) was used as a probe liquid. Measurements were performed at room temperature, using the sessile drop method. For static measurements, the Young–Laplace equation was employed to determine the contact angles by interpolating the contour side droplet, using the OCA20 instrument software. The dispensed volume for static measurements was 2 µl. Regarding dynamic measurements, the tangent method at the left and right three-phase points of the side droplet image was employed to obtain the contact angle measurements over time and the dispensed volume was 1µl.

### 2.5. Electrochemical measurements

The electrochemical performance of the nanocomposites were investigated in a 2-electrode configuration using a Swagelok Cell (SKU: ANR-B01, Singapore) and 6M KOH as the electrolyte. The hydrophobic propylene microporous membrane (Celgard 2500) was modified by a two-step process: soaking with acetone at 20 °C for 5 min followed by dipping in aqueous 6M KOH solution at 20 C for 1 h.[41] The cell was assembled by sandwiching separator-soaked-electrolytes between the carbon nanofoam grown on carbon paper. Prior to the test, electrodes and a modified separator were dipped into the electrolyte solution for several hours. Cyclic voltammograms, charge-discharge tests, and electrochemical impedance spectra were recorded using a PALMSENS4 electrochemical workstation. Before recording the original data, the prototype cells were scanned at 100 mV/s scan rate for 1000 cycles. The cyclic voltammetry at different scan rates ranging from 0.02 to 1 V/s and charge-discharge at different currents of 150 µA to 2 mA were carried out. The areal capacitance was calculated from a cyclic voltammogram using the equation: $C_{areal} = {\int I\, dV}/{A.v.V}$, where *I* is the current, *v* is the scan rate, *A* is the geometric area of the electrode, and *V* is the voltage of the device. Energy density (*E*) and power density (*P*) of devices were estimated via the relations of $E = 0.5\, C_{areal}V^2$ and $P = {(E.v)}/{V}$, respectively. For estimating volumetric capacitance and volumetric energy density, total thickness of two electrodes were considered. Single electrode capacitance = 4 × device capacitance. The volumetric capacitance of electrode materials was estimated by dividing the areal capacitance by the total height of two symmetric nanocomposite electrodes. The gravimetric capacitance of the device is equal to volumetric capacitance divided by estimated mass density. The areal capacitance of the device is also calculated from the charge-discharge profile using the relation: $C_{areal} = \frac{I_d.\int V(t).dt}{A.\int V.dV}$. The electrochemical impedance spectroscopy was conducted in the frequency range of 1 Hz to 0.1 MHz at open circuit potential with a 10 mV *a.c.* perturbation. The relaxation time constant ($\tau_0$) was estimated from the corresponding frequency ($f_0$) at -45° phase angle using the equation of $f_0 = 1/\tau_0$.

For the *in-situ* Raman spectro-electrochemical measurements, a symmetric device was fabricated with two similar electrodes separated by the electrolyte-soaked separator. A hole with a diameter of 0.1 cm was made on the top electrodes and separator such that the laser could reach the bottom electrode without any interaction with the top electrode and separator. A similar size of hole was also created on the bottom electrode to maintain the same geometrical area of both electrodes. A thin glass slide was placed on the top electrode and the full device was sealed with Kapton tape. The electrical connections between the electrodes and the potentiostat were given by platinum wires since the Cu wire was corroded by the KOH electrolyte. For this study, a Renishaw *In-Via* micro-Raman spectrometer with a diode-pumped solid-state laser (wavelength of 532 nm) and an 1800 grooves/mm grating was used. The laser was focused on the sample using a 50× long working distance objective and the incident power on the sample was 1.75 mW, below the threshold for laser-induced degradation. Each spectrum was acquired using 5 accumulations, 10 s each.

## 2. Results and Discussions

The morphology of as-grown and annealed (at increasing temperature) N-doped hydrogenated amorphous carbon nanofoam is shown in Figure 2(a-e). The porous structure can be clearly observed. No observable changes in the morphology of annealed carbon nanofoam is noticed, as compared to the pristine vacuum annealing even at high temperature of 900 °C. The nanofoam studied here is composed of a compact film and a porous film, as it can be seen from Figure 1c, where the backside view shows the porous nanofoam grown onto compact layer. The compact film provides good adhesion to the current collector (carbon paper in our case) and is efficient for the electron transport,[42] whereas the highly porous structure is responsible for the charge-storage at the electrode/electrolyte interface via double layer formation. We previously reported that[42] the electrical resistivity of this kind of nanoporous structure does not change upon bending, which is attributed to the damping of strain of porous structure and compact layer. A representative cross-sectional micrograph of CFs overlaid to elemental maps is shown in Figure 1f. Transmission electron micrographs of pristine CFs are shown in Figure 1(h-i), and annealed CFs in Figure 1(j-l). Pristine CF shows particles with an amorphous structure, with no visible graphitic planes or chains; in CF-600 and CF-900 crystalline structures are clearly distinguishable. High-resolution micrographs reveal the presence of onion-like particles (Figure 1(j-l)), which can generally be found in pulsed laser deposited carbon nanostructures.[43]

The deposition is quite uniform over the area of 2 cm × 2 cm area in a single production run. We measured the height at different positions over 2 cm and the estimated average

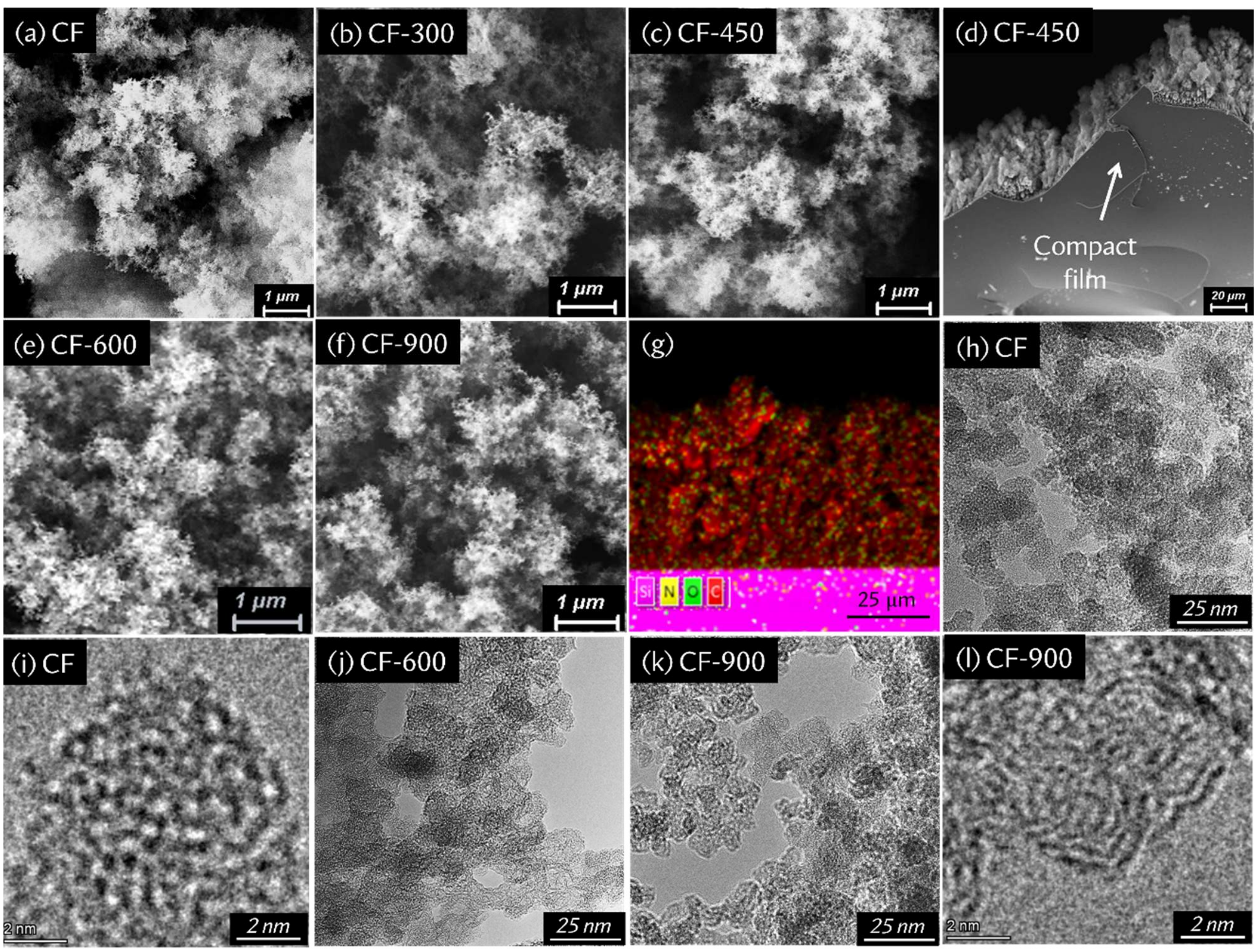

Figure 1: Scanning electron micrographs of (a) pristine N-doped hydrogenated amorphous carbon nanofoams and nanofoam vacuum annealed at (b) 300 °C, (c-*d*) 450 °C, (*e*) 600 °C, and (*f*) 900 °C. *(g)* Representative *cross-sectional* electron dispersive X-ray layered image of amorphous carbon nanofoam showing uniform distribution of nitrogen and oxygen over the carbon surface. Transmission electron micrographs of (h-i) pristine carbon nanofoams and nanofoam annealed at (j) 600 °C, and (k-l) 900 °C, highlighting the difference between the (i) amorphous and (l) crystalline structures.

thickness of pristine, CF-300, CF-450, CF-600, and CF-900 is about 62.6 (±5.9), 56.3 (±4.5), 57.6 (±8.4), 57.4 (±7.0) and 54.6 (±3.3) µm, respectively (Figure 2a). The thickness of our nanofoams is much higher than many electrode materials reported (below 10 µm or even in nanometer scale) for supercapacitor application.[23][44] The deposition area can be further scaled with our set-up by translating the substrate holder. Moreover, we have also shown in our previous study that carbon nanofoams with more than one characteristic in terms of thickness, volumetric void fractions and amount of constituent elements can also be grown from a single deposition.[33] A slight reduction in thickness at high annealing temperatures is due to the evaporation of species, which is confirmed from TGA measurement (Figure 2b). The initial weight loss up to 100 °C is attributed to the elimination of adsorbed moisture. The second weight loss observed up to 580 °C and associated to the release of $H_2O$ and $CO_X$, is attributed to the crystallization of amorphous carbon. Above 900 °C the mass of the sample decreases significantly, with evaporation of $H_2$ and CO. This result justifies our selection of annealing temperatures for the present study. Since porosity is one of the key factors for supercapacitor applications, the quantitative estimation of the porosity of our materials (volumetric void fraction, in our case), the elemental compositional ratio, and the mass density of as-grown nanofoams are evaluated by EDXS measurements coupled with the EDDIE software (the measurement details can be found in the Experimental section and our previous report[33], and the fitted spectra using the EDDIE software are provided in Figure S1).[33] By providing elemental composition details and Si substrate reference as inputs, the mass density of as-grown pristine nanofoam is estimated to be 77 mg/cm$^3$. However, once annealed under vacuum, the estimated mass density drops to ~42 mg/cm$^3$ for all annealed CFs (Figure 2a). The mass density of the nanofoam is in the range of ultralow-density materials (<300 mg/cm$^3$), and hence our material can be referred to as an ultralow-density carbon nanofoam. Using the mass density of nanofoams and comparing it with that of graphite (2.2 g/cm$^3$ as a reference), the estimated volumetric void fraction of the nanofoam is in the range of 95-98%, confirming its highly porous structure. As can be seen from the EDXS layered image (Figure 1g), oxygen and nitrogen are uniformly deposited on the carbon matrix. Thanks to the PLD approach, N-doped CFs is successfully obtained using $N_2$ background gas, unlike other techniques employing toxic and unsafe environments like $NH_3$ to grow N-doped nanocarbons. The quantitative information on the presence of carbon, oxygen, and nitrogen is estimated from EDXS and XPS measurements (Figure 2c-d). The discrepancy in the atomic concentration of elements obtained from the two techniques is due to their own limitations. EDX probes the volume of the film while XPS is more sensitive to the surface. In general, both techniques reveal the increase of the carbon content with annealing temperature whereas the content of functional groups (oxygen and nitrogen) decreases (Figure 2c-d). For CF-900, the nitrogen content is extremely low. It should be noted that the carbon nanofoams also contain hydrogen, as they were synthesized under $N_2$-$H_2$ environment (discussed later in the section on Raman spectroscopy and XPS) and due to the unwanted moisture adsorption on the surface.

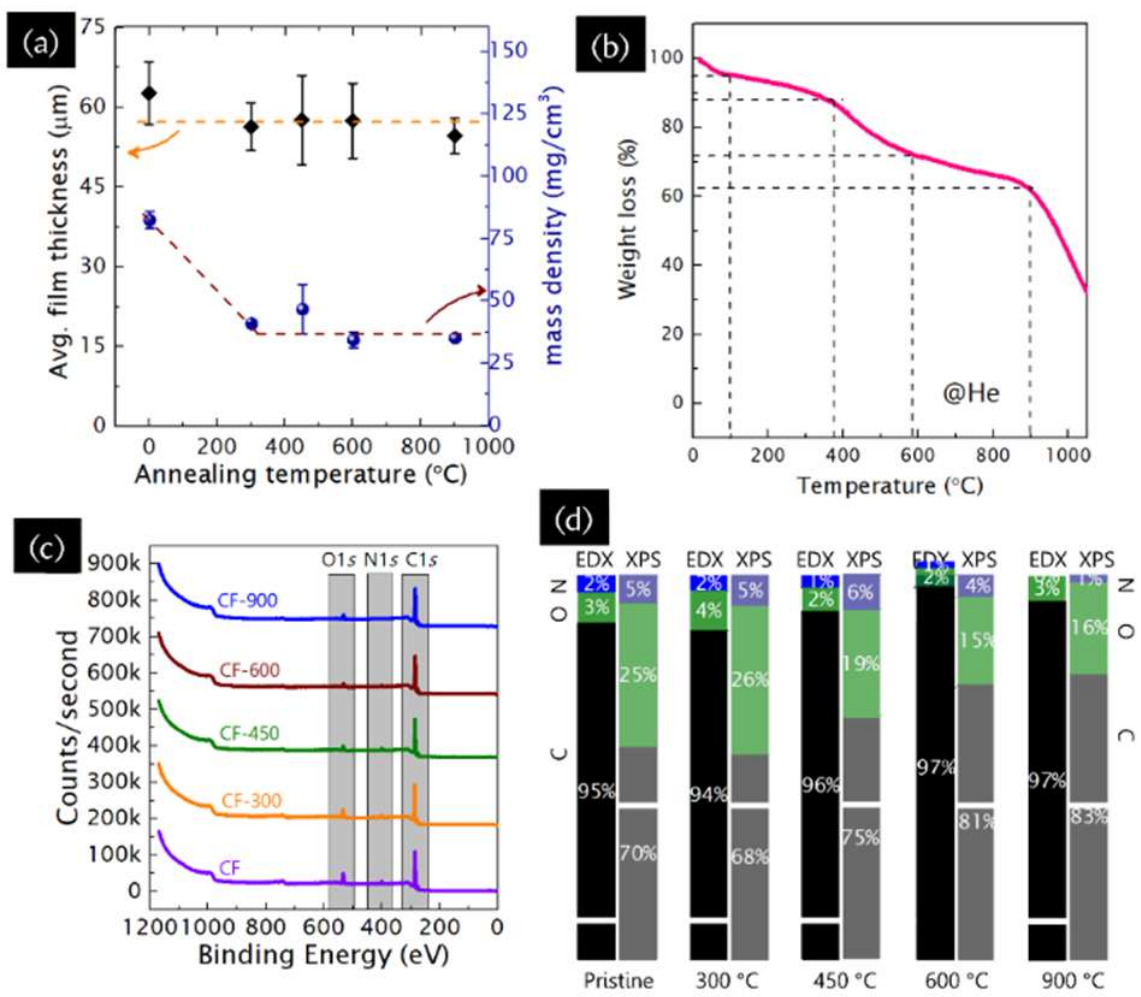

Figure 2: (a) Plot of average thickness and mass density of pristine and annealed nanofoams. (b) Thermogravimetric analysis of pristine carbon nanofoam under He gas. (c) XPS survey spectra of pristine and annealed carbon nanofoams. (d) Elemental compositional ratio of carbon, nitrogen, and oxygen of pristine and annealed nanofoams obtained from EDXS and XPS.

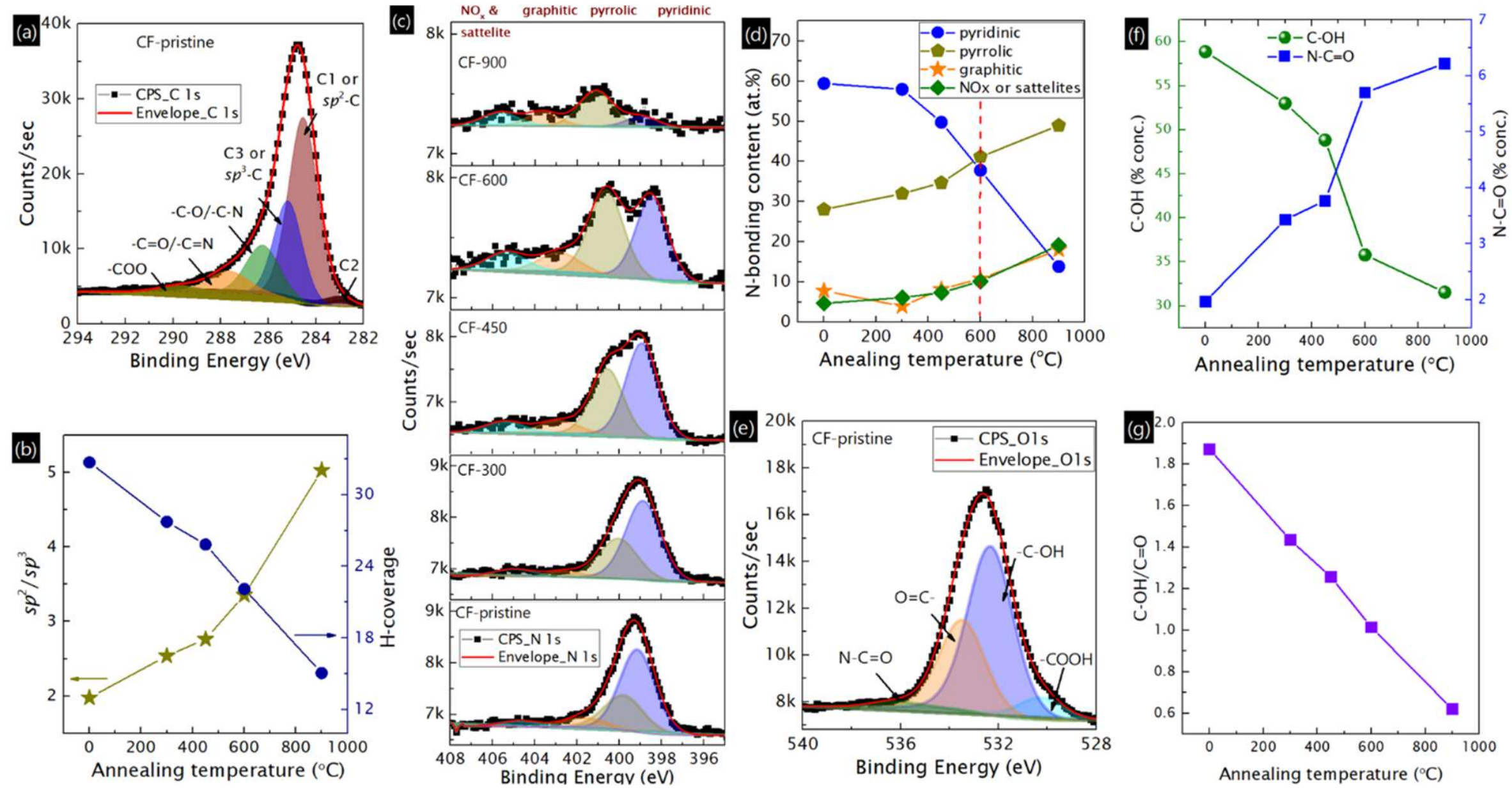

Figure 3: **XPS results**. (a) C1*s* spectrum of CF-pristine with deconvoluted peaks, (b) variation of $sp^2/sp^3$ content and H-coverage of CF with annealing temperature, (c) N1*s* spectrum with deconvoluted peaks, and (d) variation of N-bonding contents of pristine and annealed CFs. (e) O1*s* spectrum of CF-pristine with deconvoluted peaks. Variation of (f) C-OH and N-C=O content and (g) ratio of C-OH to C=O with annealing temperature

For a better understanding of the structural improvement of the nanofoam upon annealing, the core-level C1*s* spectra of as-grown and annealed carbon nanofoams are further analysed and shown in Figure 3a and Figure S2a-d of supplementary material. The C1*s* spectra consist of $sp^2$-C (at around 284.5 eV), $sp^3$-C (at around 285.2 eV), the peak at around 283 eV (C2) associated with *sp*-C and vacancy defects, and the peaks associated with functional groups in the range of 286 -290 eV (see Table S1 of supplementary material). The impact of annealing on the $sp^2/sp^3$ ratio is found to increase with respect to the annealing temperature (Figure 3b). In our case, the $sp^3$-C for pristine sample consists of $sp^3$ C-C and $sp^3$ C-H, which is difficult to distinguish from XPS. We also estimated the H-content using the equation of $H - coverage(\%) = \frac{sp^3-C \times 100}{sp^2-C+C2+sp^3-C}$.[45][46] and found that the H-content of nanofoam decreases as annealing temperature increases. Thus, the increase of $sp^2/sp^3$ ratio and reduction of H-content upon annealing (Figure 3b) is an indirect indication of structural improvement via transformation of $sp^3$-C to $sp^2$-C and $sp^3$C-H to $sp^3$ C-C in the structure. Apart from C1*s* spectra evolution, we observe a distinct evolution of N-bonds from the N1*s* spectra upon annealing of pristine nanofoams (Figure 3c). Each nitrogen feature is extracted from the N1*s* spectrum by deconvoluting it into four configurations: pyridinic-N (at around 399 eV), pyrrolic-N (at around 400 eV), graphitic-N (at around 402 eV), and $NO_x$ and satellite peaks (at around 405 eV) (Table S2 of supplementary material). Among the N-bonds, pyridine/pyrrole-N defects provide additional ion adsorption sites, improve the wettability, enhance the ion diffusion kinetics, and hence improve the overall charge-storage performance of the electrode material, whereas graphitic-N improves the electrochemical stability.[5] These N-bonds are well known to contribute pseudocapacitance, and quantum capacitance, and increase the double-layer capacitance.[47] The extracted N-bonding contributions indicate that pyridinic-N content decreases and other N-bonds increase with the increasing annealing temperature (Figure 3d). This indicates an alteration of the N-bonding configuration upon annealing. From Figure 3c, it can also be seen that the pyrrolic-N and graphitic-N are more prominent from the CF annealed at or above 450°C. For CF-600°C, a nearly equal amount of pyridinic-N and pyrrolic-N is observed (Figure 3d). At the higher temperature, pyridinic-N transforms into stable pyrrolic-N and graphitic-N. The O1*s* spectrum of pristine carbon foam with deconvoluted peaks such as COOH-group, C-OH group, C=O group, and N-C=O group is shown in Figure 3e and Figure S2(e-h) in the supplementary material (see also Table S3). Although a clear trend in COOH and -C=O content is not observed, the content of C-OH increases (Figure 3f) and an evolution of -C=O group over C-OH group is noticed with the increase in annealing temperature. More precisely, the ratio of -C-OH/O=C- decreases from 1.87 to 0.62 upon annealing and became one for CF-600 (Figure 3g). It has been reported that C-OH is influential for the improved capacitance of vertical graphene nanosheets, whereas COOH group is unstable and detrimental on the charge-storage performance.[23] On over it, N-C=O content is found to be increased upon annealing (Figure 3f). This result is also an indication of O-bonding configuration alteration upon the vacuum annealing.

Fourier Transform Infrared absorption spectroscopy is another analytical technique used to highlight the presence of functional groups within the carbon foam (Figure 4a). By analysing the FT-IR spectra collected in transmission on the carbon foam films deposited on silicon, we observe a weak band around 2150 $cm^{-1}$, which we assign to the stretching vibrations of C≡N or C≡C bonds. We also observe a significant increase in the background of the FT-IR spectra with the increasing annealing temperature. This is presumably related to scattering effects and/or increased light absorption due to the changing electronic properties of the annealed foams. Another important evidence comes from the C-H stretching region (just below 3000 $cm^{-1}$). The aliphatic C-H stretching bands (with $sp^3$ carbon) are quite intense in the pristine carbon nanofoam and the sample annealed at 300 °C, but they tend to vanish for samples annealed at higher temperatures due to the loss of hydrogenated species, which is in a good agreement with XPS results. The FT-IR observations confirm the evidence collected using Raman spectroscopy regarding the restructuring of the carbon film with increasing annealing temperature, which implies the gradual increase of the $sp^2$-content organized in honeycomb domains (as discussed below).

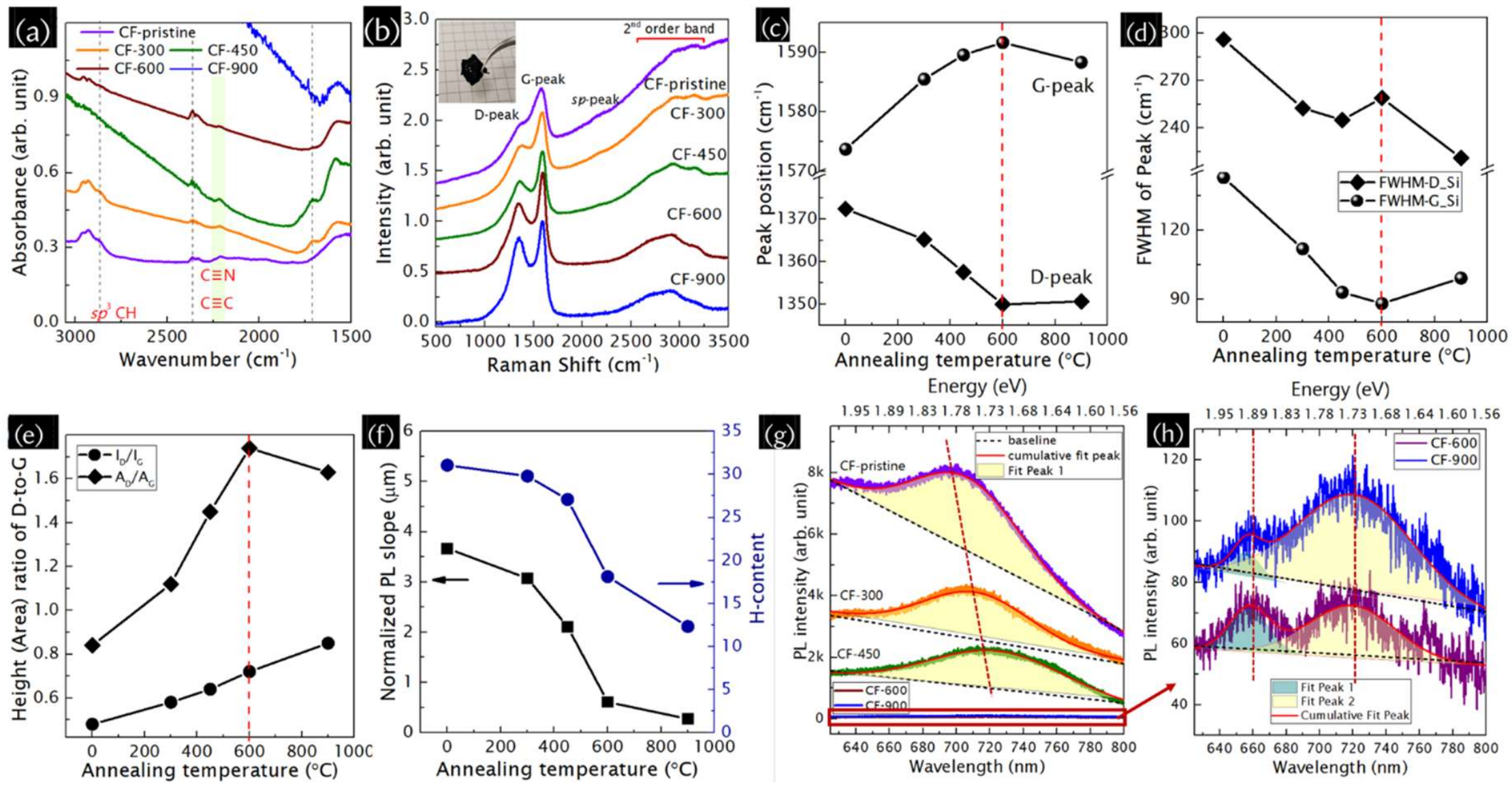

Figure 4: Evolution of (a) FT-IR spectra and (b) Raman spectra of frontside or porous-side of CF with annealing temperature. Plot of (c) position of D-peak and G-peak, (d) FWHM of D-peak and G-peak, (e) $I_D/I_G$ and $A_D/A_G$, and (f) normalized photoluminescence or PL slope and H-content with respect to the annealing temperature. For the analysis of Raman spectra in (b-e), Raman spectra were recorded in three different positions and averaged out to extract the Raman parameters. (g-h) PL spectra of pristine and annealed N-doped hydrogenated amorphous carbon nanofoam. (laser wavelength is 514.5 nm).

In order to validate the structural improvement of nanofoams upon annealing, we investigated the Raman spectra from the porous side and also the compact side.[48–50] The carbon nanofoams are free-standing (see the inset of Figure 4b), and the annealed film can be easily detached from the silicon substrate due to poor adhesion. Figure 4(b) shows the structural evolution of Raman spectra recorded from the porous side or top of the carbon foam. Typically, the Raman spectrum of carbon foam consists of D-peak and G-peak, with a very weak contribution of *sp*-peak and followed by a broad band in the high-frequency range, which is the typical spectrum of amorphous carbon.[51] To investigate the structural changes in nanofoam upon annealing, we fit the D-peak with the Lorentzian function and the asymmetric G-peak with the Breit–Wigner–Fano function. The clear observations on the annealing effect of carbon foams are the reduction in photoluminescence (PL) background, and changes in peak position, FWHM, and intensity or height (area) ratio of D-to-G peak ($I_D/I_G$ and $A_D/A_G$) (Figure 4c-f). The peak position of D-peak, FWHM of D- and G-peak, and the normalized PL slope of carbon foams decrease, whereas the G-peak position, $I_D/I_G$ and $A_D/A_G$ are found to increase upon annealing until the sample CF-600 ensures the structural ordering of nanofoam. G-peak normalized PL spectra of all N-doped hydrogenated amorphous carbon nanofoams is supplied in Figure S3a of supplementary material, which is in good support with the PL analysis described above. For CF-900, the increasing or decreasing trend of the above parameters was flipped. A closer inspection of the spectra shows a weak hump at around 1150 $cm^{-1}$ in CF-600 (an enlarged and clearer version can be seen from Figure S3b of supplementary material), which could be an indication of pentagon-heptagon structures[52] or onion-like carbon structures (as seen from TEM, Figure 1j and 1l) in the carbon foam. Since the D-peak is quite broad for all samples, the peak at 1150 $cm^{-1}$ became very weak or shadowed under the D-peak. We have also estimated the H-content from normalized PL slope employing the similar empirical formula used for hydrogenated *a*-C. [49]. We found a good agreement of H-content estimation from Raman spectra with the H-coverage estimation from XPS (Figure 3b). These changes in Raman spectra can be attributed to the $sp^3$-C to $sp^2$-C transformation, $sp^3$ C-H to $sp^3$ C-C transformation, reduction of N-content, and change in N-bonding as observed from XPS results.

Moreover, we found an interesting result on the Raman spectra recorded from the backside of the free-standing film, which is the compact film as indicated in Figure S3c of supplementary material. The backside Raman spectra are quite different from the Raman spectra acquired on the porous side. For instance, the deconvolution of first-order Raman spectra of CF-600 from compact-side resulted in D-peak at around 1362 $cm^{-1}$ with FWHM of 325 $cm^{-1}$ and G-peak at around 1570 $cm^{-1}$ with FWHM of 152 $cm^{-1}$ in contrast to those from porous-side (D-peak at around 1350 $cm^{-1}$ with FWHM of 259 $cm^{-1}$ and G-peak at around 1591 $cm^{-1}$ with FWHM of 88 $cm^{-1}$ ). The difference in the Raman spectra of carbon foams between the compact side and porous side is obvious due to the different morphology and different localized heating gradients during annealing as the compact side is supported by the substrate. Moreover, we also noticed the presence of curvature peaks[53] at around 400 and 700 $cm^{-1}$, and *sp* C≡N peak at around 2210 $cm^{-1}$ is clearly observed in both CF-450 and CF-600. However, *sp* C≡N has disappeared for CF-900, which is in good agreement with EDXS and XPS results, indicating that N-content is low. The curvature peak can be attributed to the presence of carbon-onion-like structures as reported in Ref. [53], and it is also in good agreement with TEM (Figure 1j and 1l) and the XPS peak at around 283 eV. In contrast to the frontside or porous side of foams, the noticeable changes in the backside of foams are the decrease in FWHM of D-peak from 375 to 314 $cm^{-1}$ and the shift in G-peak position from 1564 to 1587 $cm^{-1}$ (see Table S4 in supporting material). In summary, from both sides (porous and compact), a significant structural improvement has been noticed for carbon nanofoam upon annealing, and CF-600 is found to be an optimized $sp^2$-C/ $sp^3$-C ratio with appropriate amount of N-content and N-bonding.

The room temperature PL spectra of all carbon nanofoams recorded with excitation of 514.5 nm (2.41 eV) is shown in Figure S3(c) of supplementary file, consisting of Raman region and PL region. The PL region is shown in Figure 4g-h. To provide quantitative information, the PL spectra were fitted with a Gaussian lineshape for CF-pristine, CF-300 and CF-450. A broad emission at about 707 nm is observed for CF-pristine with a tail at lower energy. The PL background became lower, and the position of 707 nm emission was red shifted, in the nanofoams annealed at higher temperature. The change in PL background is in line with the estimation of PL background obtained from Raman analysis. The shift of emission peak is due to the increased amount of $sp^2$ cluster in the $sp^3$ matrix.[54] The PL intensity is highly reduced for both CF-600 and CF-900, due to the reduction of H-content in the structure as discussed before as well as N-content. It was reported for hydrogenated amorphous carbon nitride films that the PL intensity increases by incorporating more N.[55] Looking in detail, we noticed a clear splitting of broad emission into a distinct emission spectra at around 657 nm (1.89 eV) along with the emission band at 720 nm (1.73 eV) for both CF-600 and CF-900. Moreover, the FWHM of 657 nm decreased from 28 nm to 19 nm for the nanofoam when annealing temperature was raised from 600 to 900 C. The reduction in FWHM may be due to the reduction of disorder in structure. When comparing those two emissions, the intensity ratio of 657 nm to 720 nm obtained is 0.84 for CF-600 and 0.23 for CF-900. It has been proposed for carbon dots with an average size of 5 nm that the small graphitic domains (1-2 nm) surrounded by amorphous carbon core form an inter-band-energy gap of 1.69 eV and produce near-infrared PL emission.[56] As the near infra-red PL emission arises from direct recombination of carrier in small graphitic subregions, it is independent of surface states formed by surface functional groups attached on the amorphous carbons. The surface states trap the excitation in the amorphous carbon core upon absorption of light and produce visible PL. Furthermore, the increased NIR PL intensity and the decreased visible PL is due to the suppression of electron trapping in the surface states. [56]

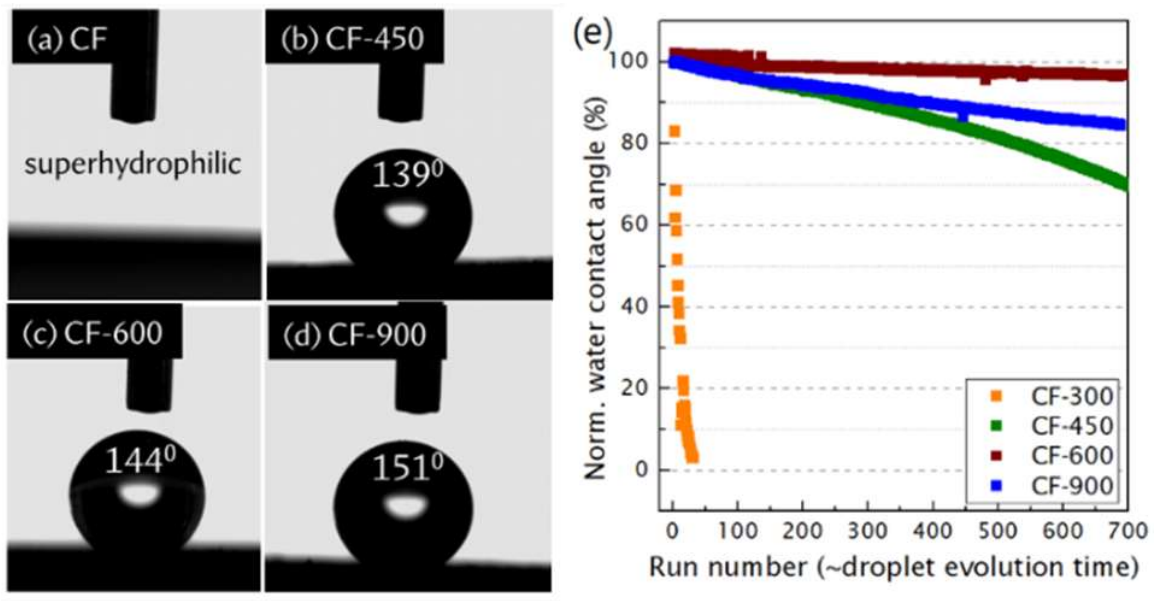

Figure 5: Static water contact angle of (a) pristine nanofoam and annealed nanofoams (b) CF-450 °C, (c) CF-600 °C, and (d) CF-900 °C. The volume of water droplet was 2 µL. (e) Changes in water contact angle of pristine and annealed nanofoams. Run number in X-axis of (e) is equivalent to the evolution time of droplet, which is around 3 min approximately.

Wettability of the electrode is one of the key factors for charge-storage performance in supercapacitors. It has been seen that both the pristine and CF-300 are superhydrophilic (Figure 5a) in nature. The superhydrophilicity can be attributed to the huge volumetric void fraction (as estimated from the morphological information), to the presence of functional groups containing hydrogen, oxygen, and nitrogen on the surface (as obtained from elemental analysis), and to the amorphous structure with huge defect sites.[7] Moreover, higher hydrogen and nitrogen content on the carbon surface, as for the case of pristine sample (Figure 3b and 4f), leads to the formation of polar interactions and hydrogen bonding between water and the nanofoam, resulting in much faster water absorption and hence the water droplet spreads spontaneously. Hence, we could not be able to record the water contact angle for the pristine sample. A similar behavior was observed for CF-300 once the water dropped on the surface, but the rate of spreading over the surface was a slightly lower compared to the pristine case, which can be seen from the plots of normalized water contact angle versus evolution of droplets (Figure 5e). The contact angle increased with annealing temperature and nanofoams annealed at 450 °C and above are found to be superhydrophobic in nature (Figure 5b-d). The contact angle of CF-450, CF-600 and CF-900 is 139° (±3°), 144° (±3°) and 151° (±3°), respectively. The water contact angle CF-450, CF-600 and CF-900 at different time intervals is provided in the Figure S4 in supplementary material. Although they are highly porous and morphologically similar to the pristine counterpart (as visible in the morphological analysis), the hydrophobicity can be attributed to the increased carbon to functional groups ratio (as shown from elemental analysis), improved structural quality (as seen from Raman and XPS analysis), and lowered hydrogen content (Figure 3b and 4f). At the same time, we noticed a decrease in water contact angle with the droplet evolution time for CF-450 and CF-900, whereas it was stable for CF-600. The decrease in contact angle for CF-450 is attributed to the higher amount of hydrogen, whereas for CF-900 is due to the more abundant oxygen species (Figure 3).

In order to study the influence of wettability and the $sp^2$-C/$sp^3$-C ratio on the charge-storage performance, cyclic voltammograms (CV) for the symmetric supercapacitor device have been performed at various scan rates ranging from 0.02 to 1 V/s (Figure 6a and Figure S5a-d of supplementary material). All CV shapes are near-rectangular and mirror-symmetric and remain unchanged at various scan rates. A comparative CV of pristine and annealed carbon foams at 0.1 V/s is shown in Figure 6b. The areal capacitance is estimated and plotted in Figure 6c as a function of the scan rates. Despite the hydrophilic nature of CF-pristine and CF-300, which is preferable to have effective electrode/electrolyte interaction while used for supercapacitor applications, the lower areal capacitance and voltage of the device indicate that wettability is not an influential factor for the charge-storage improvement. Moreover, a theoretical investigation reported in Ref. [57] suggests that the improvement in surface wettability leads to high energy density but lowers the power density of the device. Based on the investigation, the observed increase in areal capacitance for the device composed of annealed foams, which are hydrophobic, is attributed to the higher but optimized $sp^2$/$sp^3$ ratio, improved structural quality, and optimized functional groups. The CV and areal capacitance vs scan rate of CF-600R (annealed at 600 °C for a shorter time and soaked in electrolyte for a shorter time than CF-600 before device assembly) is shown in Figure S5h of Supplementary Material. Moreover, the voltage of the symmetric device increased from 0.6 V for pristine to 1.1 V for CF-900, although pristine NF contains higher $sp^3$-carbon content. Enhancing the voltage of the device simultaneously increases the energy density and the power density (Figure S5h of Supplementary Material). We have already mentioned the disappearance of the $sp^3$ C-H vibration for CF-450 onwards (as seen from FT-IR analysis), and the reduction in H-content indicating the transformation of $sp^3$ C-H to $sp^3$ C-C along with the increase of $sp^2$/$sp^3$ ratio. Moreover, $sp^3$-bonded carbon (diamond) is well known to stabilize the widened voltage.[10][11] As a result, we have seen enhanced voltage of the device with CF-900.

At the lowest scan rate (0.02 V/s), the highest areal capacitance of 58.7 mF/cm$^2$ is achieved for the CF-900 device (Figure 6c). While compared at the scan rate of 0.1 V/s, the highest areal capacitance obtained for CF-600 device is 54.1 mF/cm$^2$ (volumetric capacitance of 4.7 F/cm$^3$ and gravimetric capacitance of 137 F/g) and showed excellent retention at 1 V/s (51 mF/cm$^2$). On the other hand, the CF-900 device exhibited very poor rate performance at 1 V/s (32% against the capacitance at 0.1 V/s, Figure 6c), which could be due to a higher amount of oxygen functional groups and negligible N-bonding despite higher $sp^2$/$sp^3$ ratio compared to CF-600. It has been reported that the optimization of $sp^2$ carbon and nitrogen incorporation in doped porous carbons via thermal annealing improves the conductivity and also contributes the pseudocapacitance.[18] Moreover, CF-600 showed higher energy density at higher power density compared to all other devices, as shown in Figure 6d. The performance of the supercapacitor device and best nanofoam is compared with the existing thin-film-based supercapacitors (Table 1). Both the areal and volumetric performance of our CF devices is also compared with the device based on carbon film of thickness 400 nm[58], Mo-wire of thickness 300 nm,[59] $MoO_3$/Mo[60], laser induced graphene[61], carbon/$MnO_2$[62], CNT/polypyrrole[63], CNT/RGO[64], activated carbon fiber [65], Al electrolytic and commercial supercapacitors[66]. To check the influence of wettability on performance, we also functionalized the CF-600 using a 8M $HNO_3$ treatment (CF-600-f), and the surface was found to be superhydrophilic from the water contact angle measurement (similar to Figure 5a, and hence no image is shown here). From the cyclic voltammogram, the maximum obtained areal capacitance was 11.8 $mF/cm^2$ at 20 mV/s, and the stable operating voltage was 0.6 V only. It has been also reported that activated carbons with higher degree of functionalities exhibited the smallest anodic electrochemical stable window in organic electrolytes.[67] After functionalization, the O/C ratio, as estimated from EDXS, increased from 1.95% to 22.1% (Figure 7a). Moreover, an increase of H-content in annealed nanofoam after $HNO_3$ treatment can be associated with the change in the normalized photoluminescence slope of corresponding Raman spectra. The normalized PL slope of annealed nanofoam increased from 0.5 µm to 1.0 µm after $HNO_3$-treatment, reflecting an increase of hydrogen species in the structure. This result is in good agreement with the previous reports on the carbon structure after $HNO_3$-treatment.[68] Changes in FWHM and position of G- peak (Figure 7b) were also observed after functionalization. From FT-IR of CF-600-f (Figure 7c), a stronger C=O stretching peak relative to the C=C stretching peak was observed in nanofoam after functionalization. Thus, the additional and undesired oxygen and hydrogen functional groups were found to be detrimental as they lower the stable voltage of the device. We also examined the structure morphologically (Figure 7d-e) and found that although the CF-600-f preserved the porous structure, it agglomerated due to the *nanocarpet effect*, which is detailed in our previous report.[36] As a result, the reduced areal capacitance of the activated device is observed, since the pores of carbon nanofoams are blocked.

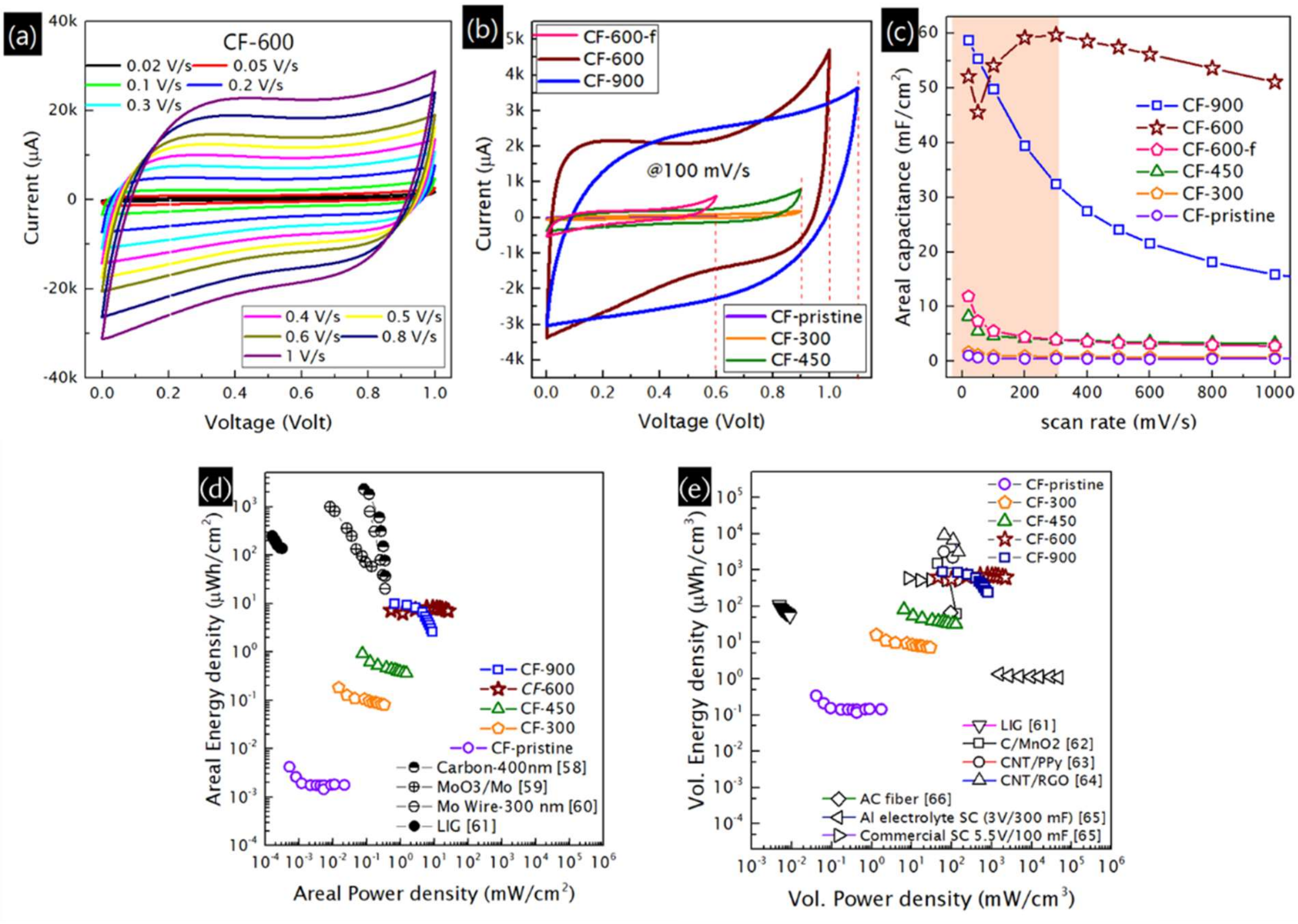

Figure 6: Supercapacitor performance of a symmetric device. (a) Cyclic voltammogram of CF-600 device at various scan rates ranging from 0.02 to 1 V/s, (b) comparative voltammograms at 0.1 V/s, (c) areal capacitance as a function of the scan rate, (d) Areal and (e) volumetric Ragone plot of CF devices and compared with the devices based on the existing literatures. In (d-e), LIG is laser-induced graphene, RGO is reduced graphene oxide, PPy is polypyrrole, and CNT is carbon nanotube.

The charge-discharge profiles of all the studied devices are shown in Figure 8a and Figure S6a-d of supplementary material. Like CV, the linear triangular and symmetric charge discharge profile is obtained for all the studied samples. The highest electrochemical stable window for pristine, CF-300, CF-450, F-600, and CF-900 is found to be 0.6, 0.9, 0.9, 1.1, and 1.2 V (Figure S6e), respectively. However, increasing the device voltage lowers the Coulombic efficiency. For instance, at the current of 1 mA, the Coulombic efficiency of the CF-900 supercapacitor with a voltage of 1.2 V dropped to 84.5% from 94% for the device with a voltage of 0.9 V. The areal capacitance of CF-300, CF-450, CF-600, and CF-900 supercapacitors were estimated to be 0.54, 2.15, 37.0, and 33.6 mF/cm$^2$ at 1 mA, respectively. Compared to the CF-900, the CF-600 device showed higher capacitance retention of 88.5% even at 10 mA (Figure 8b). To check the reproducibility, the electrochemical test of another CF-600 (CF-600R, annealed at 600 °C for shorter time and soaked in electrolyte for shorter time than CF-600) was conducted, and the estimated specific capacitance of device was 26.6 mF/cm$^2$ at 1 mA with 83.34% retention at 10 mA (Figure S6e-f of supplementary material). Figure 8c shows the rate performance and the cycle stability of the CF-600 supercapacitor, indicating its capability to retain 87% of its initial capacitance after 10,000 charge-discharge cycles. The cycle test of other samples was carried out at different currents (based on the charge-storage performance) and shown in Figure S6g of supplementary material. Lastly, at the frequency of 120 Hz, the CF-600 device also shows higher double-layer capacitance (4.7 mF/cm$^2$) compared to that of CF-900 device (1.2 mF/cm$^2$) (Figure 9a). Thus, it can be concluded that better ionic intercalation takes place in CF-600.

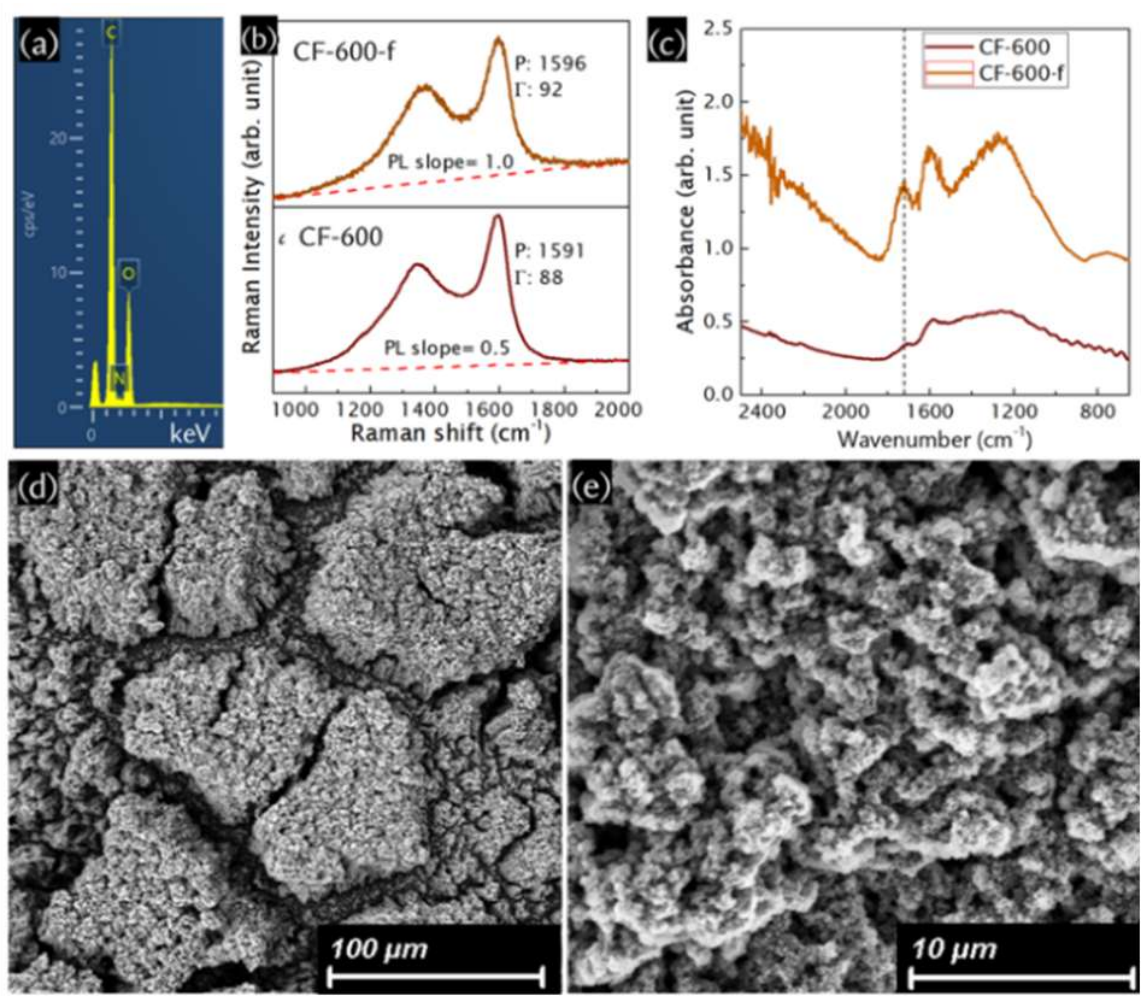

Figure 7: (a) Energy dispersive X-ray spectrum, (b) Raman spectra (acquired with 514.5 nm exciting laser), and (c) FT-IR spectra of functionalized CF-600. In (b-c), the spectra of CF-600 are provided for direct comparison after functionalization. (d-e) Morphology of functionalized CF-600 at different magnifications.

Table 1: A comparison table of binder-free thin film supercapacitor performance and single electrode performance (* represents the data obtained either from the figure using WebPlotDigitizer software authored by Ankit Rohatgi or available data from cited references. // represents asymmetric supercapacitor)

| Electrode material (Thickness) | Areal capacitance, Volumetric capacitance | Rate performance | Voltage, electrolyte used | Cycle stability |
|---|---|---|---|---|
| ***Result of supercapacitor device*** | | | | |
| Annealed N-doped carbon nanofoam–This work (57.4 μm) | 52.1 mF/cm$^2$ at 20 mV/s | 98 % at 1 V/s | 1 V, 6M KOH/modified propylene membrane | 87% after 10000 cycles at 1 mA |
| Amorphous carbon thin film [32] | 0.28 mF/cm$^2$ at 50 mV/s | 59.6%* at 1 V/s | 0.8V, 6M KOH/cellulose seperator | 99.6%* after 2000 cycles at 500 mV/s |
| superwetted vertical graphene nanosheets (415 nm) [23] | 2 mF/cm$^2$ at 100 mV/s | 99.3 % at 800 mV/s | 0.8 V, 1 M KOH gel electrolyte | 80 % after 5000 cycles at 0.1 mA/cm$^2$ |
| cumulenic *sp*-carbon atomic wires wrapped polymer (40 μm) [69] | 2.4 mF/cm$^2$ at 20 mV/s | 54 % at 1 V/s | 1 V, 6 M KOH/modified propylene membrane | 85 % after 10000 cycles at 0.1 mA |
| | 2 mF/cm$^2$ at 20 mV/s | 36 % at 1 V/s | 1 V, 1 M $Na_2SO_4$/modified propylene membrane | 54 % after 10000 cycles at 0.1 mA |
| $TiO_xN_{1-x}$ nanogrid [44] | 2.6 mF/cm$^2$ at 10 mA/cm$^2$ | - | 0.8 V, 1 M KCl | 88.9 % after 10000 cycles |
| | 0.73 mF/cm$^2$ at 10 mA/cm$^2$ | - | 2.5 V, TEA-BF4/AN | 83.9 % after 10000 cycles |
| | 0.95 mF/cm$^2$ at 10 mA/cm$^2$ | - | EMIM-TFSI | 66.2 % after 10000 cycles |
| SiC nanowires@carbon nanotubes hybrid conductive network [70] | 34.57 mF/cm$^2$ at 0.2 mA/cm$^2$ | 62.68 % at mA/cm$^2$ | 1 V, 2M KCl | 107.1 % after 10000 cycles at 10 mA |
| $MoS_2$//$V_2O_5$ (asymmetric supercapacitor) [71] | 11 mF/cm$^2$ at 0.8 mA | 36.16* % at 2 mA | 1.4 V, 1 M $Na_2SO_4$/filter paper | 80 % after 5000 cycles at 3.5 mA |
| $MnO_2$ coated vertical graphene//$Fe_3O_4$ coated vertical graphene[72] | 76 mF/cm$^2$ at 2.5 mA/cm$^2$ | 50 % at 10 mA | 2.6 V, 1M $NaClO_4$/filter paper | 79 % after 12000 cycles |
| ***Result of single electrode*** | | | | |
| N-doped annealed carbon nanofoam (57 μm) – This work | 216.4 mF/cm$^2$ at 100 mV/s | - | - | - |
| turbostratic graphene[73] | 8.2 mF/cm$^2$ at 50 mV/s | 56% at 1 V/s | 6M KOH | 120.3% after 10000 cycles at 1 mA/cm$^2$ |
| MgZn-$MnO_2$ [26] | 62 mF/cm$^2$ at 20 mV/s | 22.6% at 180 mV/s | 1.4 V vs Ag/AgCl (3M KCl), 1 M $Na_2SO_4$ | - |
| boron-doped ultranano-crystalline diamond[74] | 0.0784 mF/cm$^2$ at 20 mV/s | 16.3* % at 100 mV/s | 1V vs Ag/AgCl (3M KCl), 1 M $Na_2SO_4$ | 80% after 2000 cycles |
| boron-doped micro-crystalline diamond[74] | 0.0852 mF/cm$^2$ at 20 mV/s | 18.5* % at 100 mV/s | | |
| $MnO_2$-$V_2O_5$/$WTiO_2$ nanotube | 180 mF/cm$^2$ at 40 mV/s | 46 % at 150 mV/s | 1 V vs Ag/AgCl (3M KCl), 1.0 M LiCl | 94% (under light) & 93% (in dark) after 5000 cycles |
| Vertical graphene nanosheets (200 nm) [75] | 100 μF/cm$^2$ (6.4 F/cm$^3$) at 100 mV/s | - | 0.5 V *vs* Ag/AgCl, 1M $H_2SO_4$ | - |
| $MnO_2$ coated vertical graphene[72] | 118 mF/cm$^2$ at 10 mV/s | 29.66% at 200 mV/s | 1.4 V *vs* Ag/AgCl (3M KCl), 1M $NaClO_4$ | Stable upto 2000 cycles |
| $Fe_2O_3$ coated vertical graphene[72] | 151.11 mF/cm$^2$ at 10 mV/s | 28.24 % at 200 mV/s | 1.35 V *vs* Ag/AgCl (3M KCl), 1M $NaClO_4$ | 75.7 % after 2000 cycles |
| ZnO/Carbon nanowalls shell/core nanostructures (800 nm) [76] | 4.2 mF/cm$^2$ at 40 μA/cm$^2$ | 71.43 % at 200 μA/cm$^2$ | 0.7 V *vs* Ag/AgCl (3M KCl), 1 M KCl | Around 300% after 26000 cycles |

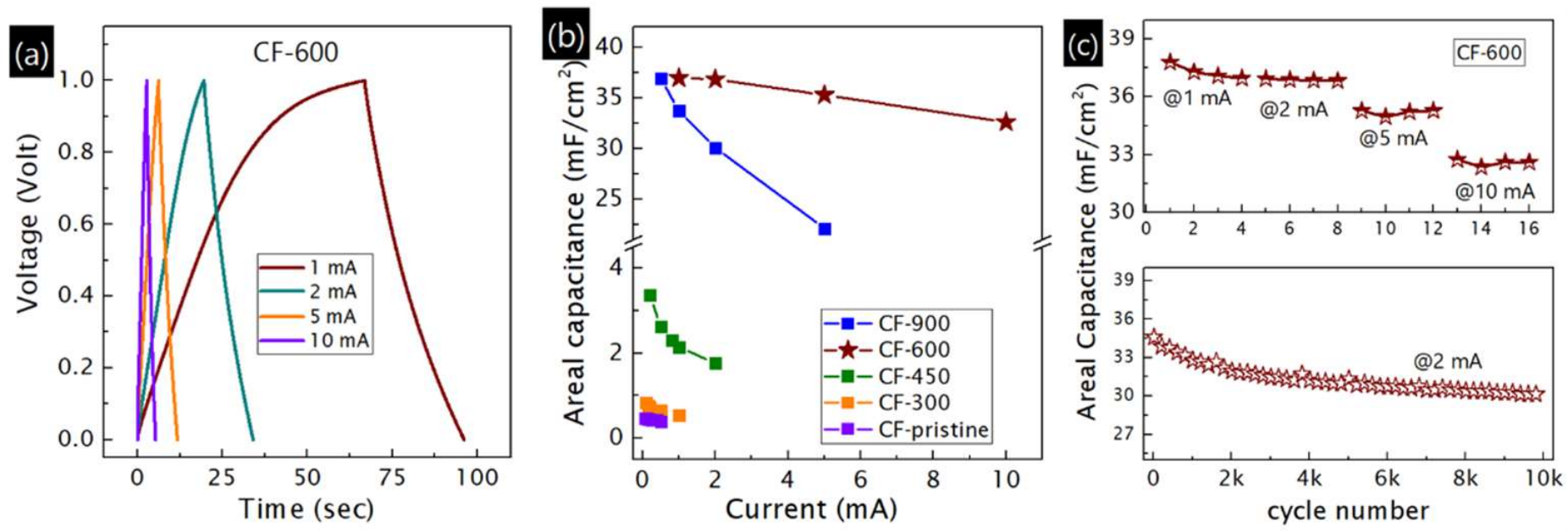

Figure 8: (a) Charge-discharge profiles of CF-600 symmetric device. (b) Estimated areal capacitance of all the studied devices with respect to the applied current. Different current was applied to the different nanofoams since they have different capacitance. (c) Rate performance at different currents and cycle stability of CF-600 symmetric supercapacitor device.

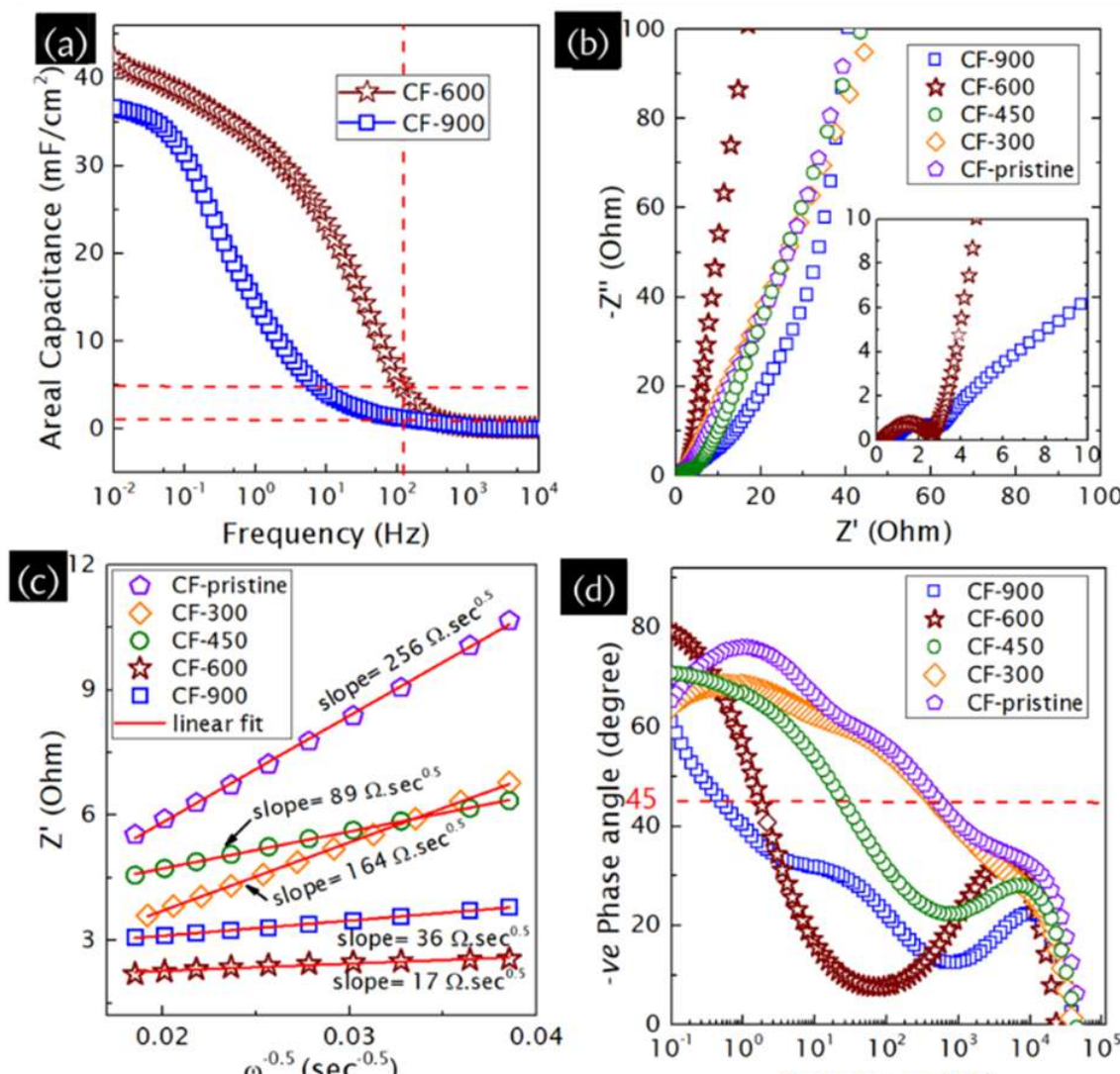

Figure 9: Electrochemical impedance spectra of carbon nanofoam devices. (a) Bode plots, (b) Nyquist plots, (c) plots of Z' vs $\omega^{-0.5}$, and (d) variation of areal capacitance with respect to the frequency.

The charge-storage kinetics of the studied devices were investigated by electrochemical impedance spectroscopy and shown in Figure 9. From the real *vs.* imaginary part of impedance, known as Nyquist plot (Figure 9b, magnified version of high-frequency region in the inset), it is confirmed that the CF-600 shows the ideal supercapacitor features compared to other devices, as the Nyquist line is more vertical in the mid-and low-frequency region. The real part of impedance (Z′) is the combined effect of ion migration, electrical conduction, and ion diffusion. For supercapacitors, ion migration takes place at low-frequency region (below 100 Hz), whereas ion diffusion occurs in the region of mid-frequency.[77] Hence, we estimated the ionic diffusion coefficient or Warburg coefficient (σ) from the slope of Z′ vs. $\omega^{-0.5}$ plot (Figure 9c).[78] The estimated σ-values of CF-pristine, CF-300, CF-450, CF-600, and CF-900 are 256 Ω/s$^{0.5}$, 164 Ω/s$^{0.5}$, 89 Ω/s$^{0.5}$, 17 Ω/s$^{0.5}$, and 36 Ω/s$^{0.5}$, respectively, indicating better ion diffusion in CF-600. The better ion diffusion is attributed to the structural and wetting properties of CF-600, as discussed before. The hydrophobic surface is well-known to repel the water and hence promote the diffusion of electrolyte ions inside the pores and reduces the interfacial resistance between the electrode surface and electrolyte [79]. Thus, better ion diffusion is obtained in CF-600, which is attributed to the structural and wetting properties. The frequency-dependent negative phase angle, known as Bode plot, of all studied devices is provided in Figure 9d, which can be categorized into the capacitive, diffusive, and resistive part in the low-frequency, mid-frequency, and high-frequency ranges, respectively. At -45°, the characteristic frequencies (time constants) of CF-600 and CF-900, obtained from the corresponding Bode plots, are 3.72 Hz (0.27 s) and 0.58 Hz (1.72 s). The higher the frequency at -45°, the faster is the charge-discharge rate of the device.

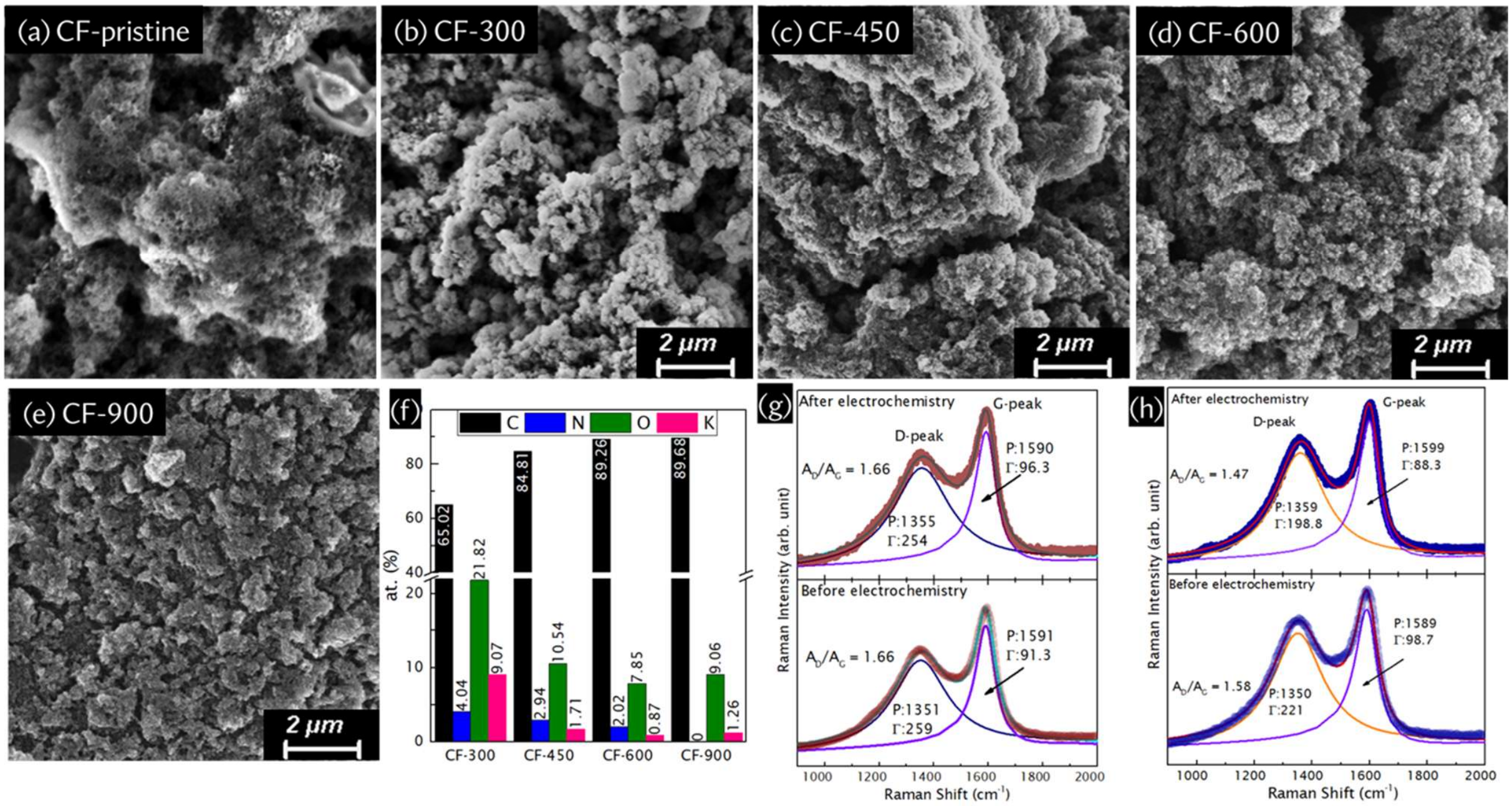

Figure 10: *Post-mortem analysis*. Scanning electron micrograph of (a) CF-pristine, (b) CF-300, (c) CF-450, (d) CF-600 and (e) CF-900 after electrochemistry. (f) Plot of atomic concentration of annealed CFs after electrochemistry obtained from EDXS measurement. Raman spectra of (g) CF-600 and (h) CF-900 before and after electrochemical investigation using 514.5 nm laser.

The initial increase of the rate performance of CF-600 device at the scan rates ranging from 0.02 V/s to 0.3 V/s, as highlighted in Figure 6c, is surprising, although all the electrodes were dipped into the electrolyte overnight and CV was carried out for several scan rates within the electrochemical stable window prior to recording the data mentioned here. Similar phenomena of increase in areal capacitance is also observed in our previous work on activated N-doped amorphous carbon.[36] To shed light on this matter, we measured the impedance spectra after cycle stability test, where we noticed a similar Nyquist plot as shown in Figure 9b but a slight deviation of the vertical line (Figure S7, supplementary material). This deviation is in good agreement with a 13% deterioration in capacitance retention after 10000 charge-discharge cycles. We also performed post-mortem analysis using SEM, EDXS and Raman spectroscopy. To do so, we disassembled the cell, and the electrode materials were washed several times with deionized water to remove the electrolyte ions residues. From the morphological investigation, we observed that the nanofoams were agglomerated while the porous structure was preserved; CF-600 showed better morphological preservation (Figure 10a-e). EDX spectra for all annealed carbon nanofoams after electrochemistry is shown in Figure S8, supplementary material.The EDXS shows a similar trend for the change in carbon, oxygen and nitrogen for carbon nanofoams with respect to the annealing temperature even after electrochemistry (Figure 10f). We note the presence of potassium, deriving from the KOH electrolyte. The analysis (Figure 10f) determined that the amount of $K^+$-ions intercalated inside the carbon nanofoams decreased from a maximum in the CF-sample, and CF-900 shows higher amount of potassium compared to CF-600. This implies that CF-pristine shows higher pore clogging, and CF-600 is effective for the $K^+$-ions insertion and de-insertion, which is in good support with the ion diffusion dynamics observed via impedance analysis (Figure 9c). We also recorded the Raman spectra for the CF-600 and CF-900 after electrochemical investigation and compared them with their pristine counterparts (Figure 10g-h). It has been seen that the D-peak position is shifted towards higher wavenumbers and the FWHM of the G-peak increases after the electrochemical process, indicating an electrochemical activation that induces structural changes in the nanofoam. The changes in Raman-extracted parameters are much higher for CF-900 after the electrochemical process. Eventually, a reduction in FWHM of D- and G-peaks, and in $A_D/A_G$ was observed, whereas the position of both peaks shifted to higher wavenumbers. The $K^+$-ions observed in EDXS are likely to have an impact on the Raman spectra too. Thus, the post-mortem investigations seem not to be an appropriate approach to understand the fact of the initial increase of specific capacitance. To deepen the understanding of the increase in areal capacitance with respect to the scan rate for CF-600 device, unlike the traditional *in-situ* spectro-electrochemistry approach with 3-electrode configuration, we fabricated 2-electrode system to record the Raman spectra of an electrode while electrochemical measurement was carried out (Figure 11a, see the details in the Experimental Section). After stability tests of the device (20 cycles CV run at 0.1 V/s), the Raman spectra of the bottom electrode (the photographic image of the electrode for *in-situ* measurements is provided in the inset of Figure 10b) were recorded after four scans at each scan rate. The CVs at different scan rates are provided in Figure S9a and the corresponding Raman spectra are given in Figure S9b of supplementary material. We have detected similar phenomena of initial increase in areal capacitance with scan rate from the *in-situ* measurements as well (Figure 11b), indicating the consistency of our result. We deconvoluted each Raman spectrum and the results in terms of peak position and FWHM of D- and G- peaks, intensity ratio ($I_D/I_G$), and area ratio ($A_D/A_G$) are provided in Figure 11c-e. The observed changes in those parameters indicate a change in graphitic quality of nanofoam.[51] Thus the changes of those parameters between the as-fabricated device and the device scanned up to certain scan rate is attributed to the electrochemical activation-induced structural changes of the carbon nanofoam.

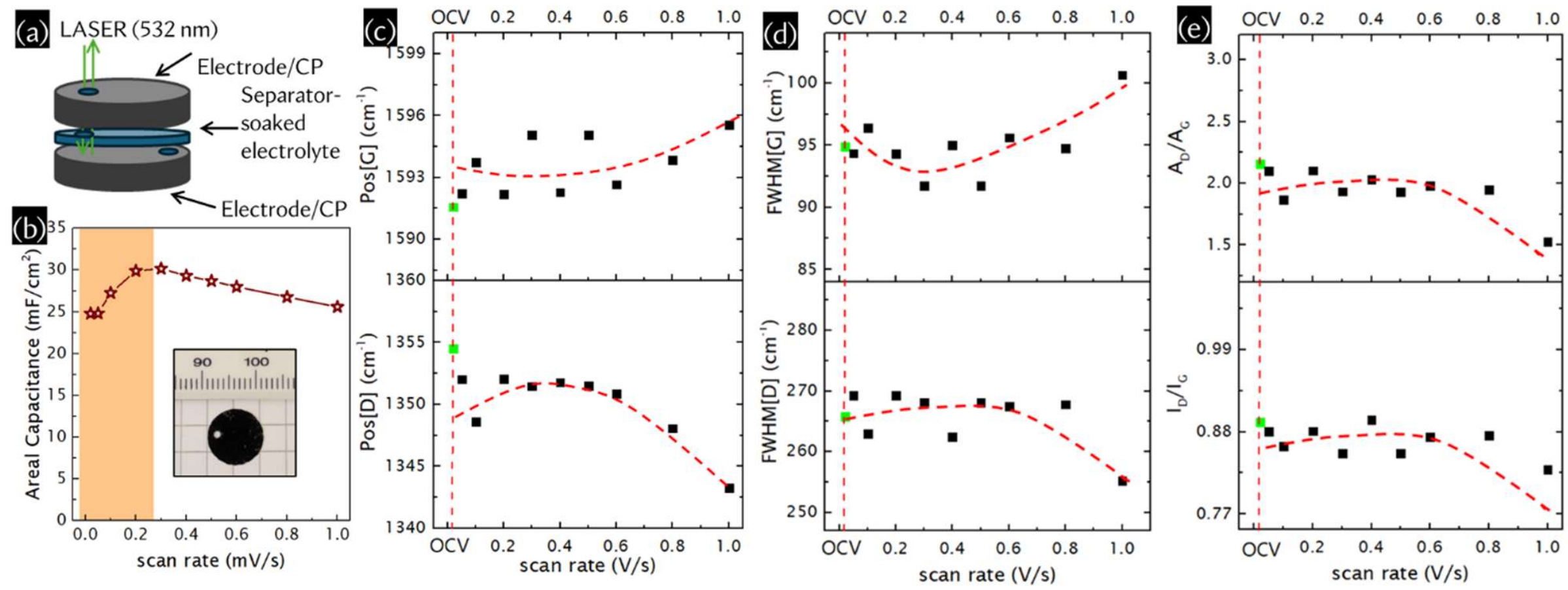

Figure 11: *in-situ* Raman spectro-electrochemistry. (a) Schematic setup of the assembled device for the *in-situ* analysis, (b) scan rate-dependent areal capacitance of the device during the *in-situ* measurement with the electrode photograph in the inset. Changes in (c) FWHM of D-peak and G-peak, (d) position of D-peak and G-peak, and (e) intensity ratio and area ratio of D-to-G peaks as a function of the scan rate. The substrate for the nanofoam here is carbon paper as it was used for electrochemical tests. In panels (c-f), the dashed red line is drawn to qualitatively show the changes (no fitting was performed) and the green square data point represent the extracted Raman parameters at the open circuit voltage (OCV) of the as-fabricated device.

## 4. Conclusion

In summary, $sp^2$-C/$sp^3$-C ratio, wettability and dopant content with specific bond configurations of binder-free and conductive-free N-doped hydrogenated amorphous carbon nanofoam were controlled by tuning the vacuum annealing temperature. The N-doped hydrogenated carbon nanofoam annealed at 600 °C showed the highest areal capacitance of 51 mF/cm$^2$ at 0.1 V/s with excellent capacitance retention and higher cell voltage compared to its pristine hydrophilic counterpart other annealed nanofoams. This result is attributed to the optimized $sp^2$/$sp^3$ ratio and an appropriate amount of N-content with specific N-bonds rather than the change in wettability. The carbon nanofoams annealed at 600 °C also preserve the porous morphology better and contain the lowest $K^+$-ion content after the electrochemical investigation and hence offer better electrolyte ionic diffusion compared to the pristine and carbon nanofoams annealed at other temperatures. The presence of $K^+$-ion content from the post-mortem analysis reveals the necessity of conducting *in-situ* study of a prototype supercapacitor device to investigate the dynamic structural changes during the electrochemical process. From the *in-situ* Raman spectroelectrochemistry, the initial capacitance enhancement of the optimized structure as a function of the scan rate is ascribed to the electrochemical activation-induced structural changes during the charge-discharge process. Finally, we emphasize such kind of ultralow dens, nanoporous and lightweight materials can be hardly achieved by other thin film deposition techniques rather than pulsed laser deposition.[80] Moreover, we anticipate a promising potential of binder-free carbon nanofoam electrodes for next-generation macro-to-micro energy storage devices.

## Authors contribution

S.G., V.R., and C. S. C. planned and conceptualized the work. S. G. and G. P. did the synthesis and characterizations of the materials. A. M. assisted in spectro-electrochemistry experiment. A. C. and G. B. did the XPS measurement. A. L. performed the FT-IR measurement, and M. T. helped in interpreting the result. R. S. did the wettability measurement. M.A. performed PL and post-mortem characterizations, G.D. and Y.P.I. carried out TEM. V.P. conducted TGA. C. C., V. R. and A. L. B. helped in data analysis and interpretation. S. G. wrote the manuscript. All authors edited the manuscript and approved the final version of the manuscript.

## Notes

The authors declare no competing financial interest.

## Acknowledgement

S.G. thank European commissions for the Marie Sklodowska-Curie Fellowship (grant no. 101067998-ENHANCER) and SEED fund from the Department of Energy, Politecnico di Milano. S.G. dedicate this research to Sir C. V. Raman. C. S. C. acknowledges partial funding from the European Research Council (ERC) under the European Union's Horizon 2020 Research and Innovation Program ERC Consolidator Grant (ERC CoG2016 EspLORE Grant Agreement 724610, website: www.esplore.polimi.it). C. S. C. also acknowledges funding by the project funded under the National Recovery and Resilience Plan (NRRP), Mission 4 Component 2 Investment 1.3 Call for Tender 1561 of 11.10.2022 of Ministero dell'Università e della Ricerca (MUR), funded by the European Union NextGenerationEU Award Project Code PE0000021, Concession Decree 1561 of 11.10.2022 adopted by Ministero dell'Università e della Ricerca (MUR), CUP D43C22003090001, Project "Network 4 Energy Sustainable Transition (NEST)". Agnieszka Jastrzebska kindly acknowledges funding from the Warsaw University of Technology within the Excellence Initiative: Research University (IDUB) programme. C. C. acknowledge support from the UKRI (Grant EP/X028844/1). We would like to thank N. G. Krishna for the useful discussion on XPS result.

**Data availability**: Data will be made available on request.

# SUPPORTING MATERIAL

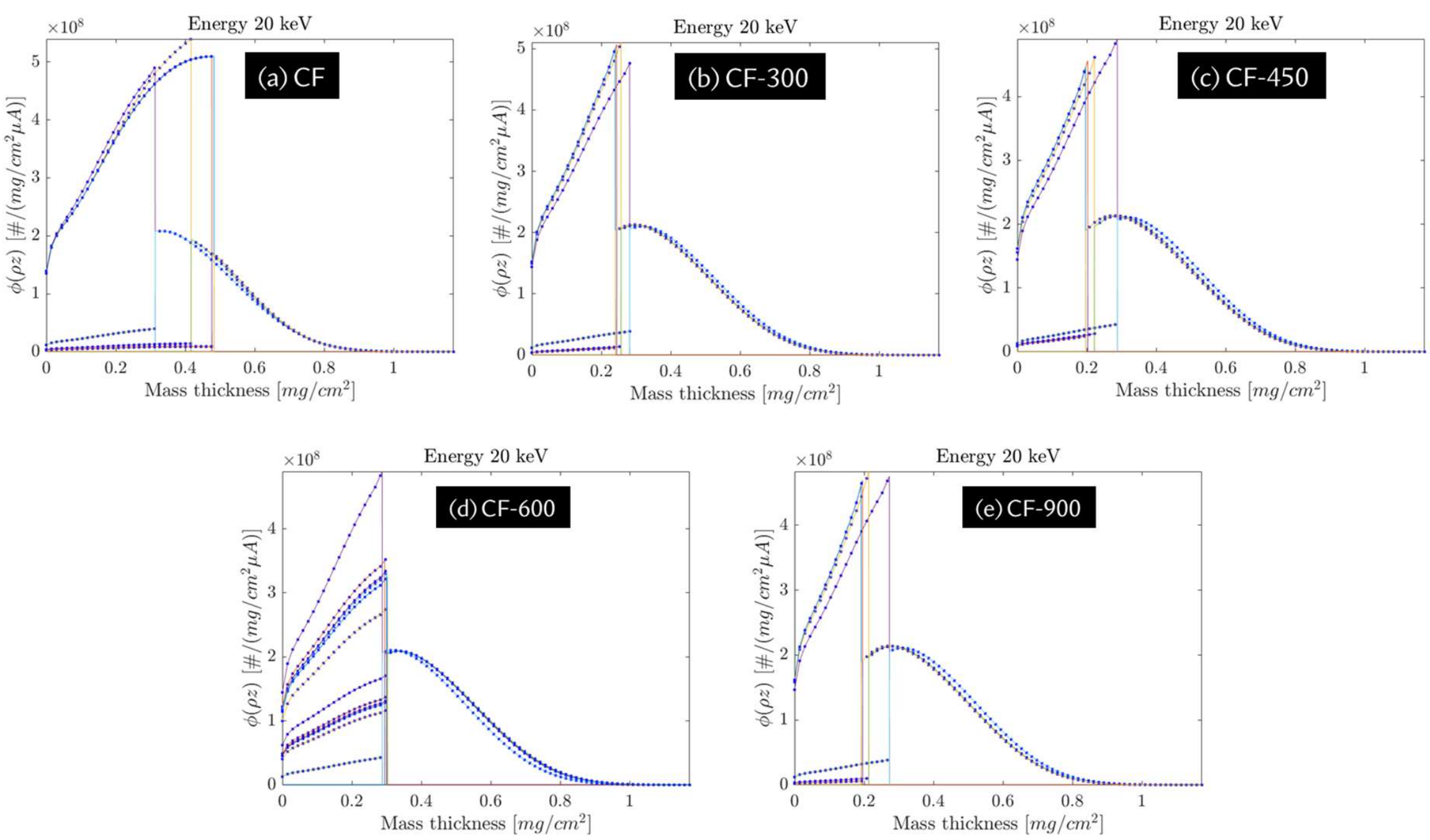

Figure S1: Simulated data of mass density calculation of (a) CF-pristine, (b) CF-300, (c) CF-450, (d) CF-600 and (e) CF-900 using EDDIE software. The y-axis of plot represents X-ray generation distribution function (ϕ(ρz)) in N-doped hydrogenated carbon nanofoams, and each coloured line corresponds to the iteration.

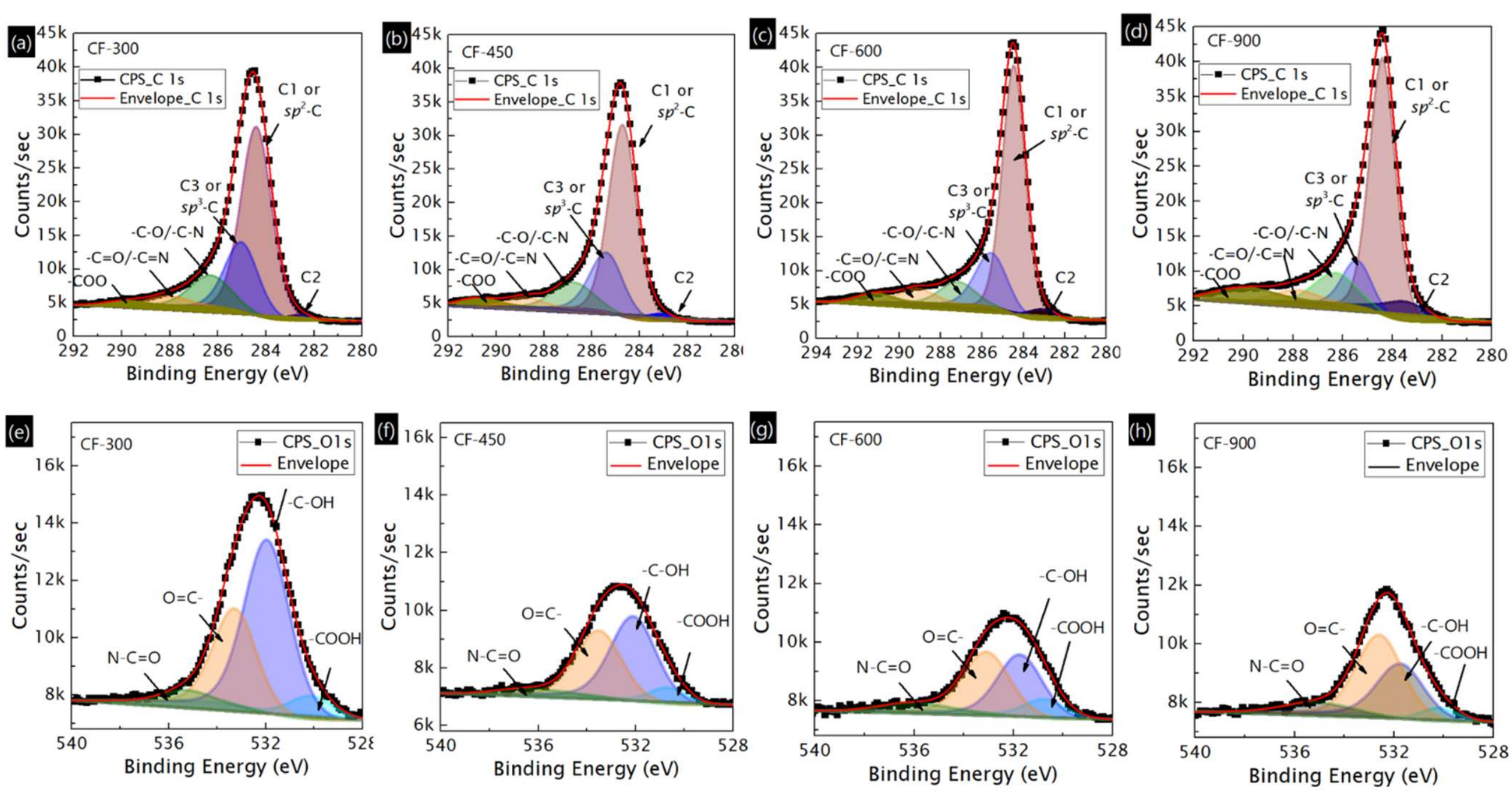

Figure S2: Top panel: XPS C1*s* spectra with deconvoluted peaks of (a) CF-300, (b) CF-450, (c) CF-600, and (d) CF-900. Bottom panel: XPS O1*s* spectra with deconvoluted peaks of (e) CF-300, (f) CF-450, (g) CF-600, and (h) CF-900.

Table S1: Extracted parameters of C1*s* deconvoluted XPS peaks of all studied carbon nanofoams.

| Sample | *sp*-C | | | $sp^2$-C | | | $sp^3$-C | | | -C-O/-C-N | | | -C=O/-C=N | | | -COO | | |
|---|---|---|---|---|---|---|---|---|---|---|---|---|---|---|---|---|---|---|
| | B.E. (eV) | FWHM (eV) | con. (%) | B.E. (eV) | FWHM (eV) | con. (%) | B.E. (eV) | FWHM (eV) | con. (%) | B.E. (eV) | FWHM (eV) | con. (%) | B.E. (eV) | FWHM (eV) | con. (%) | B.E. (eV) | FWHM (eV) | con. (%) |
| CF | 282.98 | 1.42 | 1.92 | 284.53 | 1.46 | 48.7 | 285.17 | 1.41 | 24.63 | 286.24 | 1.66 | 14.7 | 287.67 | 2 | 7.75 | 289.76 | 2 | 2.3 |
| CF-300 | 282.62 | 1.34 | 1.52 | 284.4 | 1.51 | 57.64 | 285.01 | 1.64 | 22.7 | 286.3 | 2 | 12.7 | 288.06 | 2 | 3.77 | 289.82 | 2 | 1.68 |
| CF-450 | 283.16 | 2 | 2.27 | 284.7 | 1.39 | 58.02 | 285.38 | 1.45 | 21 | 286.77 | 1.62 | 11.28 | 288.81 | 2 | 3.65 | 290.77 | 1.671 | 3.78 |
| CF-600 | 283.16 | 1.59 | 3.08 | 284.46 | 1.32 | 61.38 | 285.56 | 1.81 | 18.29 | 287.18 | 2 | 8.42 | 289.14 | 2 | 5.48 | 291.18 | 2 | 3.34 |
| CF-900 | 283.51 | 2 | 7.16 | 284.4 | 1.24 | 58.43 | 285.42 | 1.37 | 11.62 | 286.28 | 2 | 11.43 | 287.86 | 2 | 4.31 | 289.98 | 2.95 | 7.05 |

Table S2: Extracted parameters of N1*s* deconvoluted XPS peaks of all studied carbon nanofoams.

| Sample | Pyridinic-N | | | Pyridinic-N | | | Pyridinic-N | | | NOx or satellite peak | | |
|---|---|---|---|---|---|---|---|---|---|---|---|---|
| | B.E. (eV) | FWHM (eV) | con. (%) | B.E. (eV) | FWHM (eV) | con. (%) | B.E. (eV) | FWHM (eV) | con. (%) | B.E. (eV) | FWHM (eV) | con. (%) |
| CF | 399.14 | 1.95 | 59.43 | 399.79 | 2.2 | 28.06 | 401.68 | 2.2 | 7.84 | 404.69 | 2.2 | 4.68 |
| CF-300 | 398.87 | 1.95 | 57.96 | 400.03 | 2.15 | 31.99 | 402.61 | 1.91 | 3.95 | 404.99 | 2.2 | 6.1 |
| CF-450 | 398.93 | 1.8 | 49.76 | 400.56 | 1.8 | 34.71 | 402.86 | 2.2 | 8.2 | 405.45 | 2.2 | 7.34 |
| CF-600 | 398.46 | 1.8 | 37.8 | 400.61 | 1.9 | 41.17 | 402.86 | 2.2 | 10.84 | 405.36 | 2.2 | 10.18 |
| CF-900 | 398.94 | 1.8 | 13.85 | 401.1 | 1.9 | 48.96 | 403.62 | 1.8 | 18.02 | 405.67 | 1.99 | 19.17 |

Table S3: Extracted parameters of O1*s* deconvoluted XPS peaks of all carbon nanofoams studied.

| Sample | -COOH | | | -C-OH | | | O=C- | | | N-C=O | | |
|---|---|---|---|---|---|---|---|---|---|---|---|---|
| | B.E. (eV) | FWHM (eV) | con. (%) | B.E. (eV) | FWHM (eV) | con. (%) | B.E. (eV) | FWHM (eV) | con. (%) | B.E. (eV) | FWHM (eV) | con. (%) |
| CF | 530.28 | 2.05 | 7.69 | 532.32 | 2.2 | 58.88 | 533.48 | 2.2 | 31.46 | 535.31 | 2.2 | 1.97 |
| CF-300 | 530.14 | 2.2 | 7.72 | 531.94 | 2.2 | 57.83 | 533.27 | 2.08 | 30.74 | 535.33 | 2.2 | 3.71 |
| CF-450 | 530.72 | 1.8 | 8.53 | 532.10 | 2.2 | 48.84 | 533.54 | 2.2 | 38.86 | 536.46 | 2.2 | 3.77 |
| CF-600 | 530.92 | 1.86 | 23.22 | 532.10 | 1.87 | 35.79 | 533.32 | 2.2 | 35.29 | 536.45 | 2.2 | 5.70 |
| CF-900 | 530.17 | 1.97 | 11.35 | 531.86 | 2.2 | 31.54 | 532.62 | 2.2 | 50.89 | 535.44 | 2.2 | 6.22 |

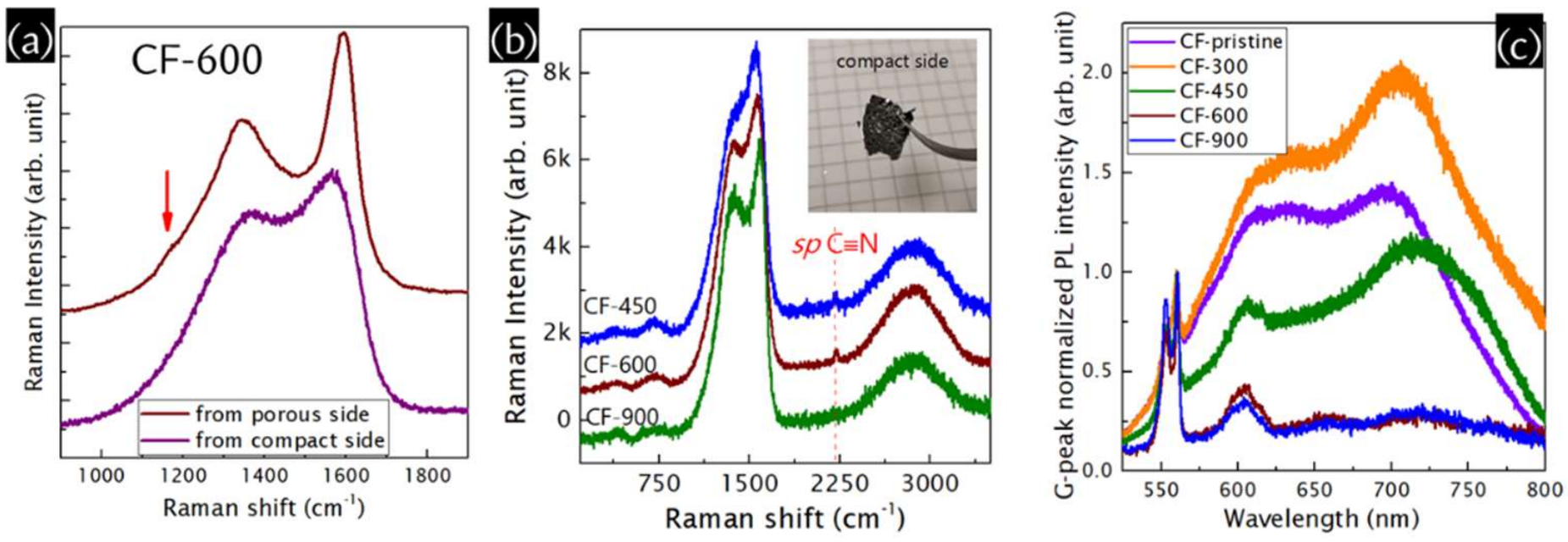

Figure S3: Raman spectra of (a) CF-600 recorded from compact side and compared with porous side, and (b) compact side of CF-450, CF-600, and CF-900 with the photographic image of film taken from compact side. (c) Photoluminescence spectra of all carbon nanofoams. Laser wavelength is 514.5 nm.

Table S4: Extracted Raman spectroscopic data of CF-450, CF-600, and CF-900 from the compact side.

| | | CF-450_compact side | CF-600_compact side | CF-900_compact side |
|---|---|---|---|---|
| D-peak | position ($cm^{-1}$) | 1374 | 1362 | 1372 |
| | FWHM ($cm^{-1}$) | 375 | 325 | 314 |
| G-peak | position ($cm^{-1}$) | 1565 | 1570 | 1587 |
| | FWHM ($cm^{-1}$) | 151 | 152 | 123 |

Table S5: Extracted PL data of N-doped hydrogenated carbon nanofoams (Lineshape is fitted with Gaussian).

| | Peak 1 | | Peak 2 | | Height ratio of peak 1-to-peak2 |
|---|---|---|---|---|---|
| | Position (nm) | FWHM (nm) | Position (nm) | FWHM (nm) | |
| CF-pristine | - | - | 707 | 75 | - |
| CF-300 | - | - | 712 | 73 | - |
| CF-450 | - | - | 724 | 94 | - |
| CF-600 | 658 | 28 | 719 | 61 | 0.83 |
| CF-900 | 656 | 19 | 722 | 82 | 0.24 |

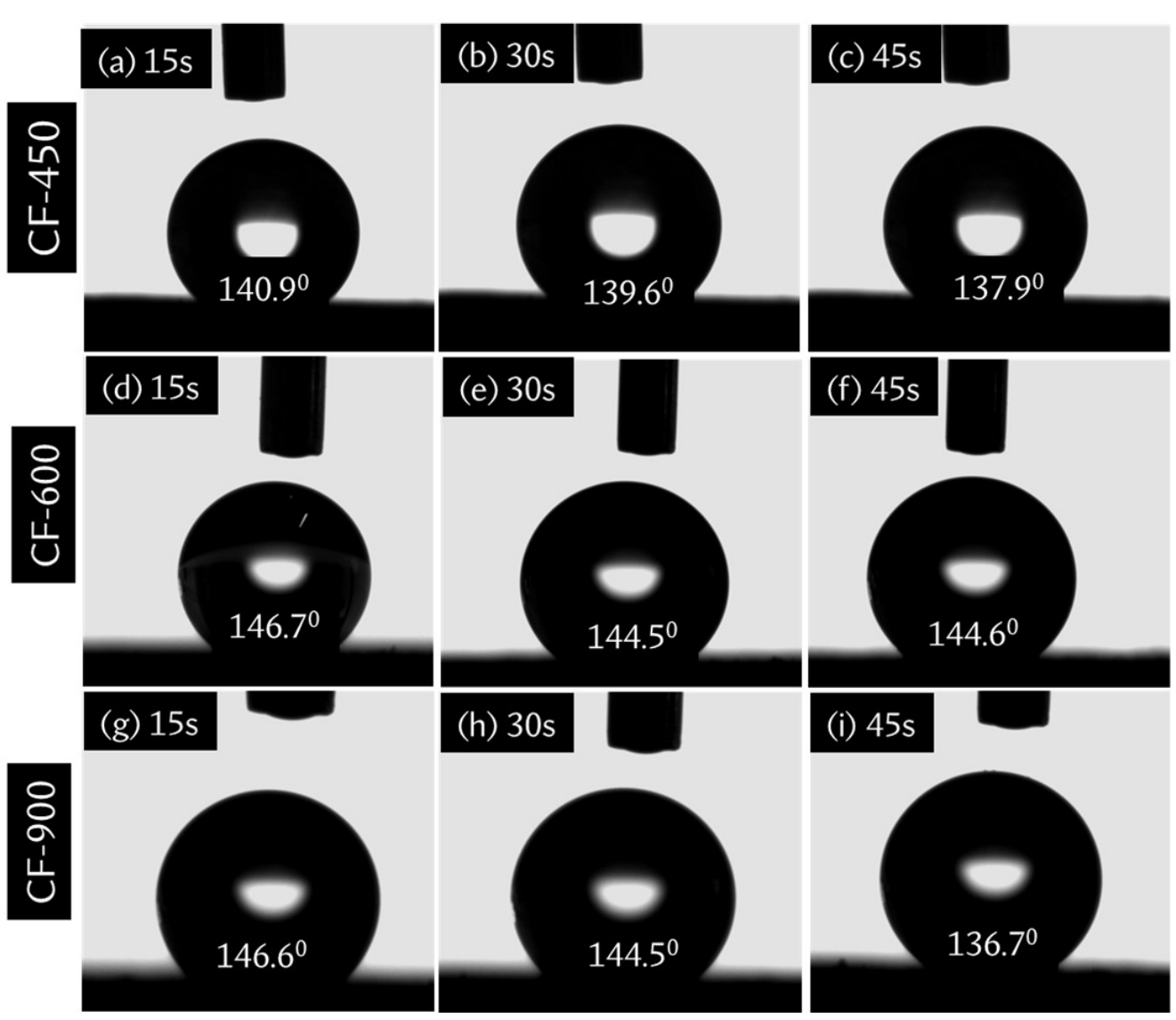

Figure S4: Water contact angle of CF-450, CF-600 and CF-900 at different time intervals.

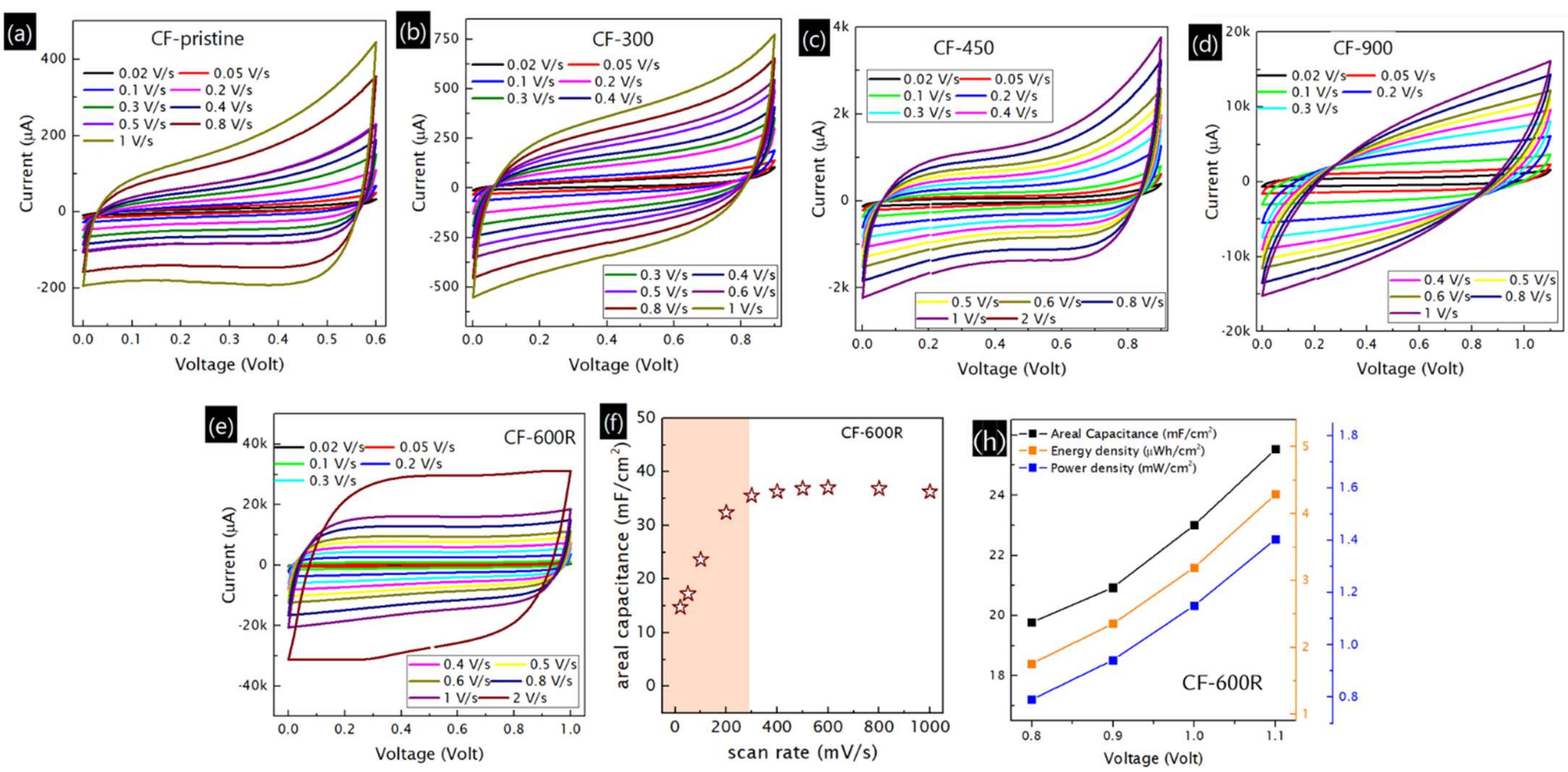

Figure S5: Cyclic voltammograms of symmetric devices made up with (a) CF-pristine, (b) CF-300, (c) CF-450, and (d) CF-900 at different scan rates ranging from 0.02 V/s to 1 V/s. (e) Cyclic voltammogram at different scan rates, (f) areal capacitance vs scan rate, and (h) the plot of the areal capacitance, energy density, and power density as a function of the voltage at the scan rate of 0.1 V/s of CF-600R supercapacitor device.

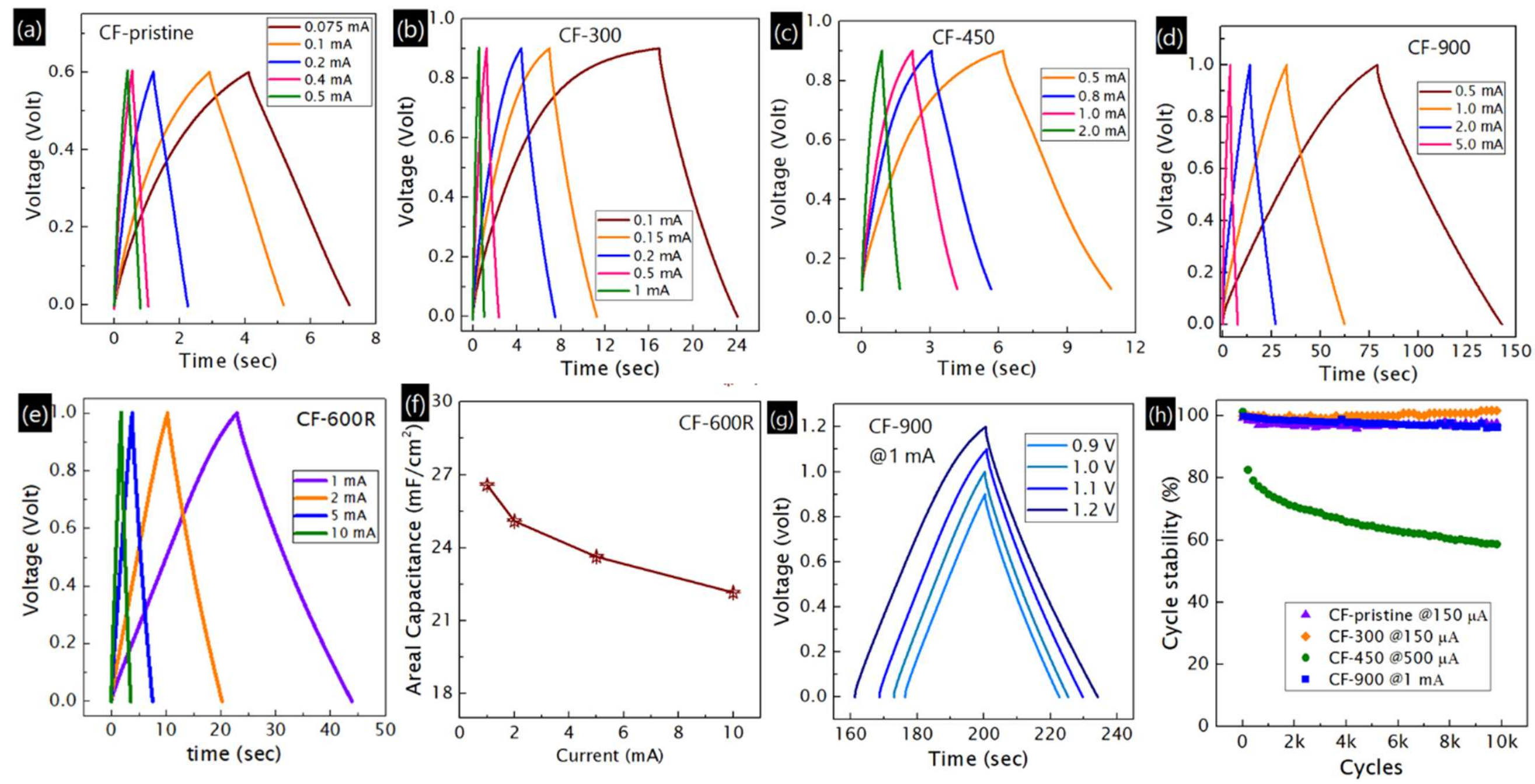

Figure S6: Charge-discharge profiles of aqueous symmetric device made with (a) CF-pristine, (b) CF-300, (c) CF-450, and (d) CF-900 at different currents. (e) Charge-discharge profile of CF-600R device at different applied current. (f) Estimated areal capacitance of CF-600R device as a function of applied charge/discharge current. (g) Charge-discharge profile of CF-900 at different voltage. (h) Cycle stability of CF-pristine, CF-300, CF-450 and CF-900.

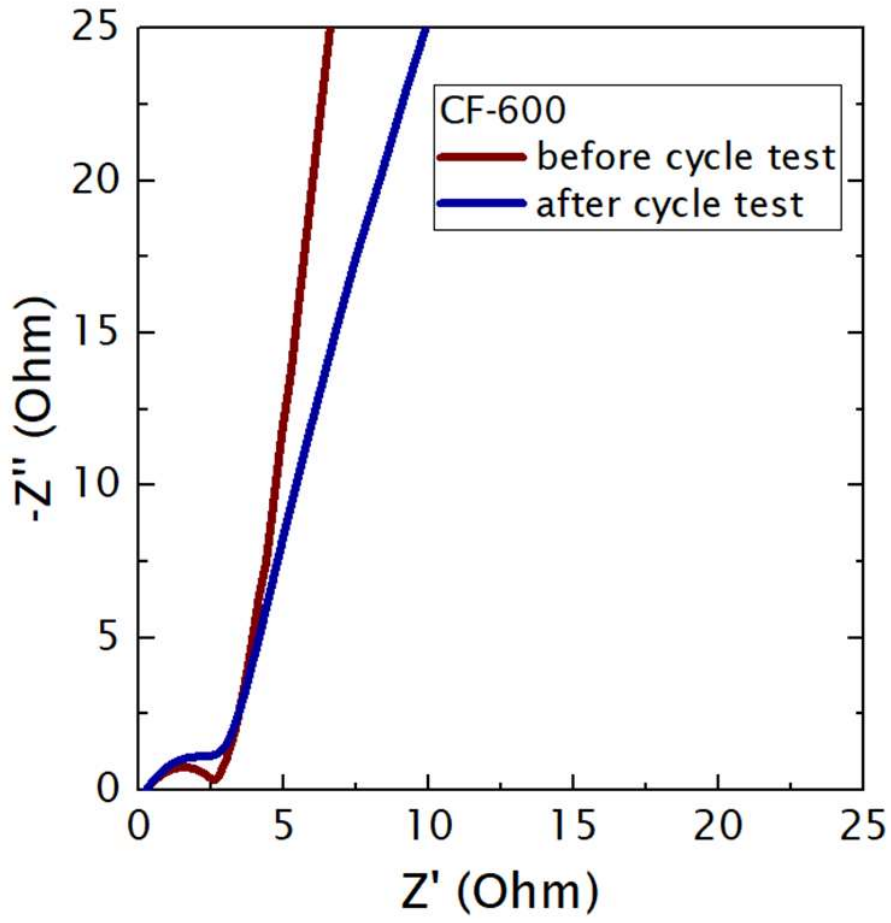

Figure S7: Nyquist plot of CF-600 before and after the electrochemical test.

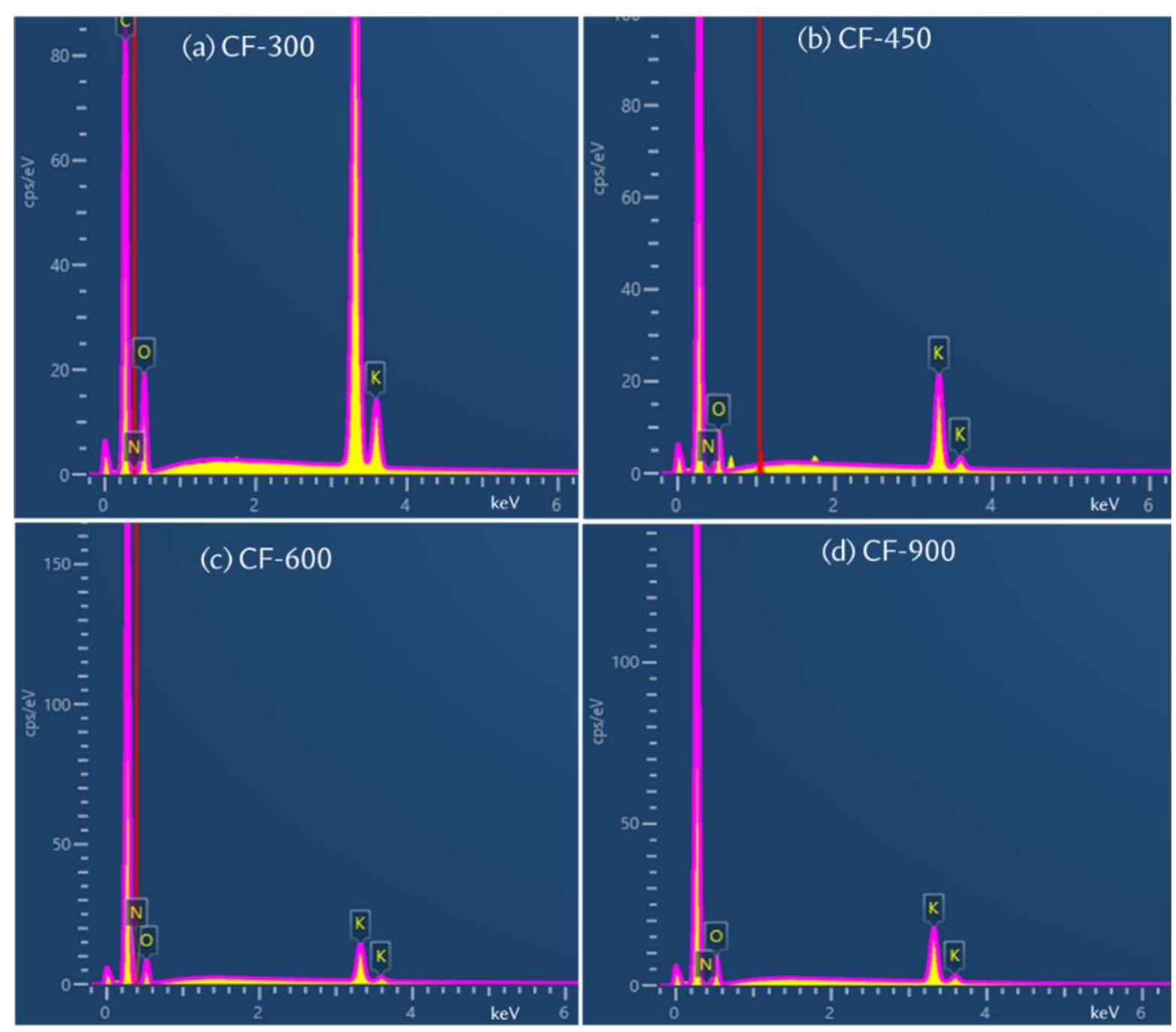

Figure S8: EDX spectra of annealed N-doped hydrogenated amorphous carbon nanofoam annealed at the temperature of (a) 300 °C, (b) 300 °C, (c) 600 °C and (d) 900 °C. Substrate is carbon paper.

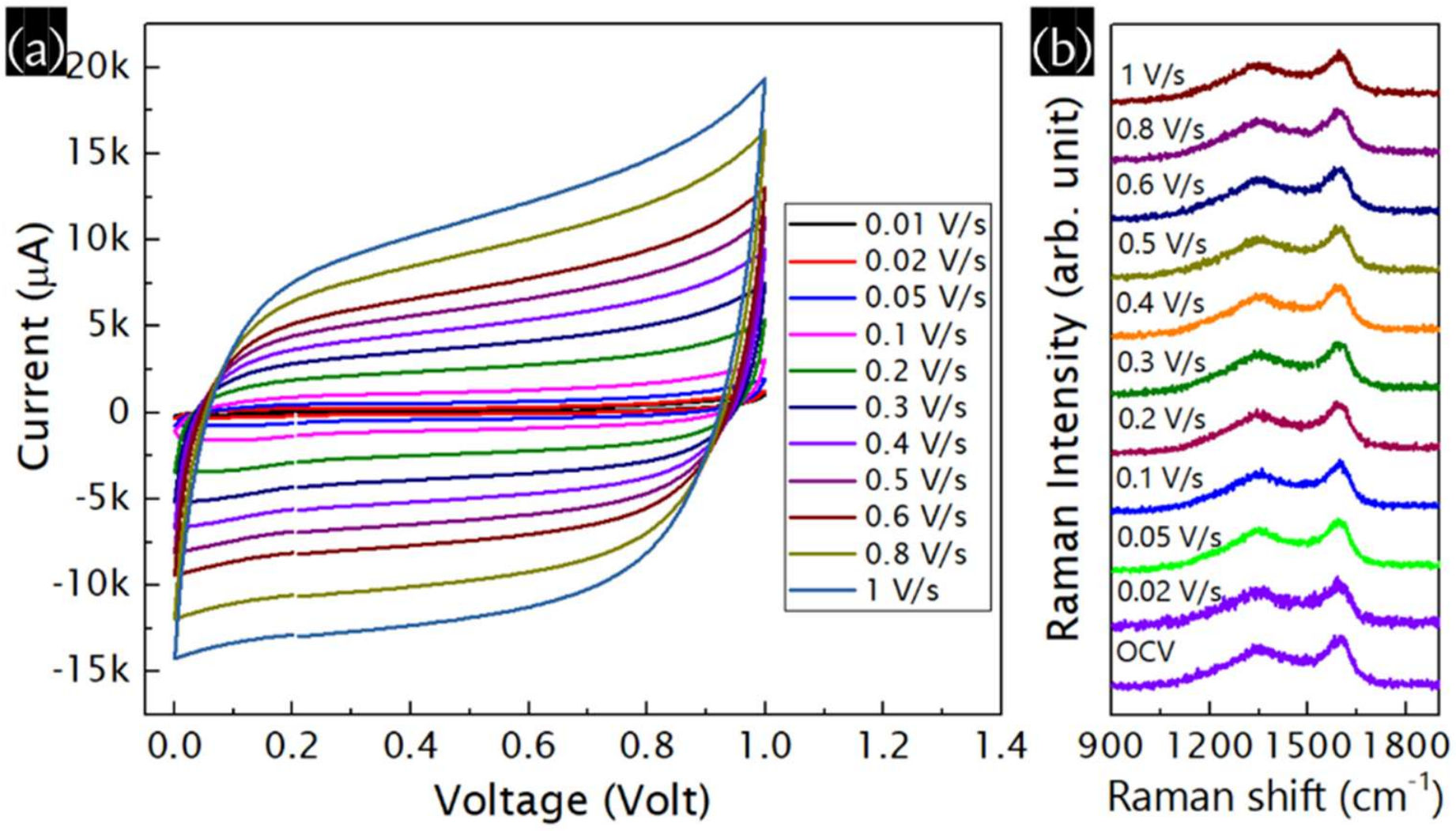

Figure S9: *In-situ* spectro-electrochemical results. The laser used to record the Raman spectra here was 532 nm and the laser power was 1.7 mW. (a) Cyclic voltammogram of the assembled device at different scan rates. (b) Normalized Raman spectra of bottom electrode at open circuit voltage and after each scan rate from 0.02 V/s to 1 V/s. Raman spectra were recorded after scanning the cyclic voltammetry four times at the specific scan rate.